\documentclass[a4paper, 12pt]{article}
\usepackage[font=bf,labelsep=space]{caption} %bold title
\usepackage{ragged2e}
\usepackage{indentfirst}
\usepackage{bookmark}
\usepackage{amsmath, amsthm, amssymb}
\usepackage{mathtools}
\usepackage{multirow}
\usepackage{rotating}
\usepackage{graphicx} %%colocar figura no texto
\usepackage{float} %%colocar figura no texto
\usepackage{titlesec}
\usepackage{anysize}
\usepackage{hanging}
\usepackage{array}
\usepackage[section]{placeins}
\usepackage{chngpage}
\usepackage{lipsum}
\usepackage{booktabs}
\usepackage{capt-of}
\usepackage{caption}
\usepackage{fancyhdr}
\usepackage[english]{babel}
\usepackage{anysize} 
\marginsize{1in}{1in}{1in}{1in}
\usepackage{ae,aeguill} % fixing figures
\usepackage{float} % fixing figures
\usepackage{natbib} % Change the styles
\usepackage{dcolumn} % with Stargazer
\usepackage{tikz} % Graphs with R
\usepackage{subfig}
\usepackage{soul} 
\usepackage{graphicx}%\usepackage{caption} % To exclude figure label
\usepackage{titlesec}
\titleformat{\section}{\singlespacing\normalfont\sc\Large\bf}{\thesection}{0.5em}{}
\titleformat{\subsection}{\singlespacing\normalfont\sc\normalsize\bf}{\thesubsection}{0.5em}{}
\usepackage{lmodern}
\usepackage{authblk}
\usepackage{enumitem}
\usepackage{setspace}
\usepackage{lscape}
\usepackage{threeparttable}
\usepackage{threeparttablex}
\usepackage{subfiles}
\setlist[enumerate]{leftmargin=2.5cm}
\setlist{nosep}

\usepackage[font=small,labelfont=bf,
justification=justified,
format=plain]{caption}

\usepackage{hyperref}
\hypersetup{
    unicode=false, % non-Latin characters in Acrobat's bookmarks
    pdftoolbar=true, % show Acrobat's toolbar?
    pdfmenubar=true, % show Acrobat's menu?
    pdffitwindow=true, % page fit to window when opened
    pdfnewwindow=true, % links in new window
    colorlinks=true, % false: boxed links; true: colored links
    linkcolor=blue, % color of internal links
    citecolor=blue, % color of links to bibliography
    filecolor=magenta, % color of file links
    urlcolor=blue % color of external links
}

\newcounter{numA}
\title{The relationship between offline partisan geographical segregation and online partisan segregation}

\author[1,3]{\normalsize Megan A. Brown}
\author[2,3 ]{\normalsize Tiago Ventura}
\author[3]{\normalsize Joshua A. Tucker}
\author[3]{\normalsize Jonathan Nagler}

\affil[1]{\small School of Information, University of Michigan}
\affil[2]{\small McCourt School of Public Policy, Georgetown University}
\affil[3]{\small Center for Social Media and Politics (CSMaP), New York University}

\date{} % clear date

\vspace{-2cm}

\begin{document}
    
\maketitle
\vspace{-1.5cm}
\begin{abstract}
\singlespacing
Social media is often blamed for the creation of echo chambers. 
However, these claims fail to consider the prevalence of \textit{offline} echo chambers resulting from high levels of partisan segregation in the United States.
Our article empirically assesses these online versus offline dynamics by linking a novel dataset of voters' offline partisan segregation extracted from publicly available voter files for 180 million US voters with their online network segregation on Twitter.
We investigate offline and online partisan segregation using measures of geographical and network isolation of every matched voter-twitter user to their co-partisans online and offline.  
Our results show that while social media users tend to form politically homogeneous online networks, these levels of partisan sorting are significantly lower than those found in offline settings. Notably, Democrats are more isolated than Republicans in both settings, and only older Republicans exhibit higher online than offline segregation.
Our results contribute to the emerging literature on political communication and the homophily of online networks, providing novel evidence on partisan sorting both online and offline. 
\end{abstract}

\thispagestyle{empty} %removes page number
    
    \pagebreak
    
    \cleardoublepage
    
    \setcounter{page}{1} 

    \doublespacing
    \justify
    \setlength {\parindent}{20pt}
    \setlength{\parskip}{4pt}

\section*{Significance Statement}
Online echo chambers are widely attributed to cause polarization and democratic malaises in American politics. Yet, Americans’ lives are shaped by strong partisan segregation offline. By linking the online networks of Twitter users to the offline residential environments of 1 million U.S. voters, we compare partisan sorting in both domains. We find that although social media users do form politically homogeneous networks, these levels of segregation are lower than those found in offline contexts. Democrats are more politically segregated than Republicans, and only older Republicans experience greater segregation online than offline. These findings challenge conventional narratives about social media as the primary driver of echo chambers and highlight the enduring role of offline environments in shaping political attitudes and behaviors.

\section*{Introduction}

The consolidation of social media and new digital technologies in today's information environment has been widely cited in both popular \citep{haidt2022yes, elbermay, hooton2016social} and academic debates \citep{sunstein2018republic, flaxman2016, bail2022breaking, settle2018frenemies} as a critical driver of the recent rise in political polarization among the American electorate.
The ``echo chambers'' narrative is often invoked as the primary mechanism by which social media causes polarization \citep{sunstein2018republic, pariser2011filter, conover2011political}. 
In this narrative, social media has facilitated the formation of overly homogenous online communities--``echo chambers''--in which citizens more easily isolate themselves from people from opposing parties and that hold distinct views about the world, consequently reinforcing their system of beliefs and reducing cross-cutting interactions with others who hold views different from the individual's own views. 

Yet, a growing body of research relying on a variety of data sources has provided a more nuanced view of this narrative, showing that the prevalence of social media echo chambers has been largely overstated \citep{eady2019many, guess2019less, gentzkow2011ideological,barbera2015tweeting}. 
Although some studies suggest that the idea of online echo chambers may be a myth, the impact of social media use on polarization remains a persistent puzzle, with some research showing that social media increases polarization, some showing social media decreases polarization, and other research showing social media has no effect on polarization \citep{allcott2024effects, allcott2020welfare, bail2018exposure, settle2018frenemies, lelkes2017hostile}, with one recent study showing that moving people out of echo chambers on Facebook had no impact on political polarization \citep{nyhan_like-minded_2023}.
To understand this puzzle, recent research has advanced an alternative causal mechanism in which social media effects must be considered conditional on users' offline experiences and context. 
Social media might actually increase contact with outgroup users and force users to engage with content that they would hardly be exposed to outside of social media, thereby increasing \citep{tornberg2022digital, bail2022breaking}.
This argument resonates strongly in the context of the United States, where geographical partisan segregation and fragmented media environments are particularly high \citep{brown2021measurement, muise2022quantifying}, and people may experience far more outgroup contact online than they do offline. There is scant research on social media and its effects on political attitudes that takes into account users' offline experiences, such as the high levels of partisan segregation in Americans' offline lives, and the degree to which  what users experience offline conditions  their online experiences.

Our paper fills this gap by measuring how Americans' online communities differ from or resemble their offline local communities. To answer this question, we develop novel individual-level measures of offline and online partisan sorting using a dataset of 950,000  Twitter \footnote{At the time of data collection, the platform was still called Twitter. We use Twitter throughout the paper.} users matched to their voter files. 
For offline partisan segregation, we use voter file information of our matched Twitter users to measure the physical proximity between every matched user and their co-partisan nearest neighbors.\footnote{Partisanship is recorded at the time voters register in 30 states and in Washington, D.C. For other states, we use imputed data estimated using procedures developed by \citep{brown2021measurement}}
For online partisan segregation, we collect the Twitter networks of each matched user in our panel and estimate the ideology of each user that our matched voter follows %friends  
using item-response estimates developed by \cite{barbera2015birds}.\footnote{This method and related techniques have been widely used to measure ideology using social media data \citep{mosleh2022measuring,lai2024estimating, eady2019many, aruguete2021news}.}
Across both measures, we consider voters to be segregated when the composition of their offline and online networks is highly homogenous, containing a high share of co-partisan voters/social media users. 
We refer to this measure of partisan segregation as online (offline) isolation, indicating that citizens in their online bubbles (local neighborhoods) are unlikely to interact with and be exposed to out-partisans. 

We first find that for both Republicans and Democrats, partisan segregation is higher offline than online, indicating that online environments actually increase exposure to the outgroup. 
Second, we find that in both online and offline environments, isolation is higher for Democrats than Republicans, meaning that in both online and offline environments, Democrats are exposed to fewer outpartisans relative to Republicans. 
Third,  this is true for almost every demographic subgroup we examine, except for older republicans that show higher offline isolation than online.
Lastly, isolation between offline and online environments is  positively correlated.

\section*{Materials and Methods}

Our article relies on two primary data sources: public voter registration files for over 180 million US voters -- which we refer to in the article as \textit{offline data}; and Twitter profiles and their followers' networks collected through Twitter APIs -- which we name in the article as \textit{online data}. The voter file data contains voter registration information up to 2021, and the online network data collection occurred from January to December 2022. 
In this section, we describe each data source in turn, giving special attention to our measures for partisan identity and how we define offline and online segregation measures. We then discuss the research procedures used to connect both data sources before concluding by addressing the ethics of using the data sources together.

\subsection*{Offline Data}

For offline data, we use the L2 voter file containing voter registration and demographic information for 180 million U.S. voters.\footnote{L2 is a private firm that aggregates individual state voter files, and appends commercially available data to individual records as well as community level publicly available data.} 
From the L2 voter file, we extract a voter's partisanship, demographic characteristics, and home address, the last of which we use to estimate partisan segregation. 
For each voter, we use the geographic location of their residential address. 
We measure the spatial distance between a voter and every other voter in the voter's state using haversine distance.\footnote{This measure is similar to Euclidean distance, but measures distance along an arc rather than directly from point-to-point to account for the curvature of Earth.}
Then, using the K-nearest neighbors algorithm,  we identify a voter's thousand nearest neighbors and consider their 1,000 nearest as voter $i$'s offline network. Using these networks, we construct an individual-level measure of offline segregation, estimating the proportion of their nearest neighbors who are from the same partisan group as the matched voters, as in \cite{brown2021measurement}.
To account for missing partisanship in the voter file, we use Bayesian process imputation, also as in \cite{brown2021measurement} . 
We provide a full description of the imputation process, auxiliary data, and accuracy metrics in the supplemental materials (SM Section 5). 

More specifically, using the voter's residential location, their thousand nearest neighbors, and their partisanship, we measure an individual's partisan segregation by calculating the proportion of co-partisans living among the voter's nearest neighbors. We call this measure \textit{offline isolation}. When isolation from co-partisan voters is low, we consider voters to live in non-sorted offline environments with high exposure to outgroup voters; when ingroup isolation is higher, partisans live in highly homogenous communities, surrounded by their co-partisans.  This indicates that a Democrat (Republican) with high \textit{offline isolation} lives in a more partisan segregated spatial area and is hardly likely to be exposed to a Republican (Democrat) in their offline geographical environment. The following equation is used to measure offline isolation: 

\[
\text{Offline Isolation} = \frac{\sum_{k=1}^{1000} \mathbb{P} \mathbb(p_k=p_i)}{\sum_{k=1}^{1000}}
\]

\noindent where $k$ is a given neighbor, $i$ is the matched voter, $p_k$ is the partisanship of the neighbor, $p_i$ is the partisanship of the matched voter.
For Democrats, the ingroup partisans are registered and imputed Democrats; for Republicans registered and imputed Republicans. 
To incorporate uncertainty from the imputation process, $\mathbb{P}$ indicates the probability that $p_k$ is equal to $p_i$. When the voter's partisanship is known directly from the voter file, this probability equals one and is equal to the posterior probability from the imputation when voters are independents or non-partisans. We use this quantity as our measure of offline partisan segregation.

\subsection*{Validating measures of offline segregation}

One potential concern with using the above measure for \textit{offline isolation} is that an individual's neighbors may not reflect the individuals they are typically exposed to at work, school, or among their acquaintances, and, therefore, this measure may not accurately reflect the ideological composition of their offline {\it networks}. 
To assess this potential concern, we carry out an additional validation analysis  that relies on survey data, asking respondents to estimate the number of their neighbors, acquaintances/friends, and coworkers that are Democrats or Republicans. The survey data was collected from an online panel with 2,270 respondents administered by YouGov from February 7, 2023 until February 14, 2023.
We then compare measures of segregation for an individuals' networks of their neighbors, acquaintances, and coworkers.\footnote{Additional survey details are available in the supporting information (SM Section 1).} If these measures are highly correlated then we would conclude that our measure of offline segregation is a reliable proxy for the level of segregation in an  individuals' daily interactions, not just their level of neighborhood segregation.
 
\begin{figure}[!htpb]
            \begin{tabular}{cc}
                \includegraphics[width=.5\linewidth]{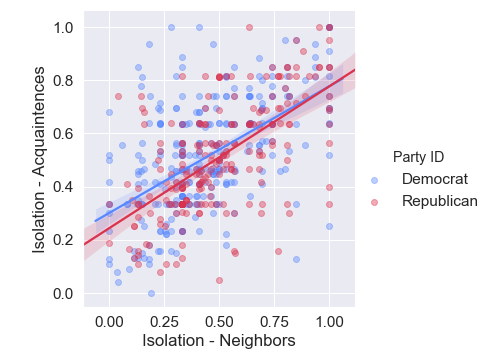}  &
                \includegraphics[width=.5\linewidth]{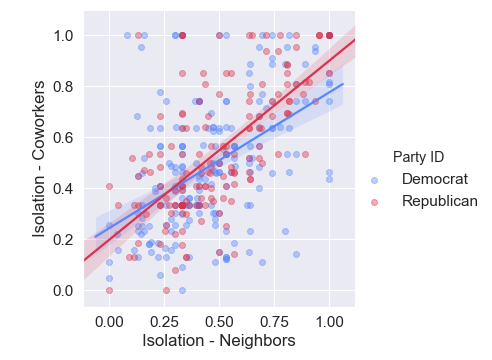} \\
                a) Acquaintances & b) Coworkers
            \end{tabular}
            \caption{Correlation between ingroup isolation in an individual's neighborhood and ingroup isolation amongst their acquaintances (left) and their coworkers (right), separately for Democrats (blue) and Republicans (red).}
            \label{fig:corr}
        \end{figure}

In Figure~\ref{fig:corr}, we show the relationship between offline isolation measured via neighbor networks (x-axis), and offline isolation  measured via acquaintance networks (left plot) and measured via coworker networks (right plot) for our YouGov survey respondents. 
We present results separately by whether the respondents identify themselves as Republicans (red) or Democrats (blue). 
We find that ingroup isolation  measured via network composition of neighbors strongly correlates to network composition measured by  both acquaintances and coworkers, indicating that our measure of partisan segregation using the voter file is a reasonable proxy for offline isolation across an  extended set of offline interactions. 

\subsection*{Online Data}
For online data, we turn to Twitter. In the last decade, Twitter has become a critical platform for disseminating and consuming political content, with a large number of voters and politicians using it as a prominent source of information \citep{barbera2019leads,  huszar2022algorithmic}. This is the case, particularly in the United States. While Twitter's user base has changed more recently, and its prominence for political information has potentially declined after its recent leadership change \citep{cava2023drivers}, our data was collected in 2021, before these changes.  We collect data from two sources: (1) Twitter's Decahose API (application programming interface), a ten percent real-time sample of all tweets, and (2) Twitter's REST API, an API that allows researchers to query both network data for a given user and posts and interactions for a given user.  We outline these sources in more detail in the following sections. 

\subsubsection*{Using Twitter's Decahose to Link Offline Data with Online Data}

To match voters with their online profiles, we use data collected from Twitter's Decahose API, a 10\% sample of all tweets collected between January 1, 2022 and December 31, 2022. 
These tweets contain users' metadata, including name and self-reported location, which we use for matching. 
We adopt the method described in \citet{hughes2021using} to link individuals in the voter file with individuals on Twitter. 
For the voter file, we take the set of voters with a unique first name, last name, and their location at the city-state level. 
For Twitter, we follow the same approach, selecting a set of users with unique first names, last names, and locations.\footnote{Twitter's enterprise APIs, including the Decahose, provide derived location information, standardizing location names. For example, if a user had ``Manhattan, NY" in their self-reported location field, the derived location would be New York, New York, USA. We use this standardized location field to match to the voter file.}
From these two datasets, we merge on first name, last name, and location.
Similar to \citet{hughes2021using}, we drop any Twitter users that are linked to more than one voter registration and any voters linked to more than one Twitter account, as there are no clear ways to disambiguate true from false positives at scale in this matching procedure. 
This results in nearly one million Twitter accounts linked to the L2 voter file. 

In Table~\ref{tab:tab1}, we show descriptive statistics of our sample of Twitter users compared to a survey of Twitter users conducted by the Pew Research Center's National Public Opinion Reference Survey \citep{pew_npors2021} and the voting population from the 2020 American National Election Survey \citep{ANES2021}. Compared to the general population,  our sample is considerably younger, more male, and more Democratic-leaning than the general U.S. population. These features of our data partially reflect the composition of Twitter users in the United States, as can be seen in Pew Research Center's National Public Opinion Reference Survey. When compared to Pew Research Data, we find our Twitter panel to have a higher share of men, be slightly more Democrat, and be slightly older than the overall Twitter user base. Overall, our matching procedure slightly accentuates the differences between the U.S. population and the Twitter user base.

% code: 06_analysis/descriptive_tables.R
%\setstretch{1.5}
\begin{table}[htpb]
\caption{\label{tab:tab1}Comparing Demographics:Twitter-L2 Panel vs. Pew Research NPORS 2021 vs ANES 2020}
\centering
\resizebox{\linewidth}{!}{
\begin{tabular}[t]{>{\raggedright\arraybackslash}p{7cm}lllrrrr}
\toprule
\multicolumn{1}{c}{ } & \multicolumn{3}{c}{Sample Values} & \multicolumn{2}{c}{$\Delta$ Twitter Panel} & \multicolumn{2}{c}{Z-Scores} \\
\cmidrule(l{3pt}r{3pt}){2-4} \cmidrule(l{3pt}r{3pt}){5-6} \cmidrule(l{3pt}r{3pt}){7-8}
Variable & Twitter Panel & Pew &  ANES & Pew &  ANES & Pew & ANES\\
\midrule
Age (Mean years) & 41.85(0.02) & 39.05(0.76) & 48.37(0.34) & -2.80 & 6.51 & -3.6600 & 19.1200\\
Gender (\% Female) & 37.77(0.05) & 47.12(2.48) & 51.55(0.78) & 9.35 & 13.78 & 3.7652 & 17.5853\\
Ethnicity (\% White) & 66.22(0.05) & 66.25(2.54) & 65.84(0.69) & 0.04 & -0.37 & 0.0146 & -0.5344\\
Partisanship (\% Democrats) & 47.29(0.05) & 41.29(2.46) & 35.35(0.76) & -6.00 & -11.95 & -2.4371 & -15.7054\\
\bottomrule
\bottomrule
\end{tabular}}
\end{table}
\doublespacing

\subsubsection*{Augmenting Twitter Data with Friend Networks}

Using all the matched accounts linked to the voter file, we queried Twitter's REST API for the friends (i.e. the people the user follows) of each matched user in the dataset. 
More specifically, in February 2023, we collected the list of accounts each of the matched users follows. 
To compare a user's offline outgroup exposure, we construct an equivalent measure of online segregation using the users' friends' networks.

For each profile in the matched user network, we measure their partisan identity using the ideal-points method outlined in \citet{barbera2015birds}. \citet{barbera2015birds} uses a homophily-based ideology measure, measuring ideology with a Bayesian Spatial Following model that treats ideology as a latent variable inferred by the list of politicians and political elites that a given user follows. 
We follow the simplified version of the method in \citet{barbera2015tweeting}, which instead uses correspondence analysis to estimate ideology, approximating the results of the full Bayesian estimation with fewer required computing resources.
This model results in a scalar value for each individual, ranging from -3 to +4, where a more negative score indicates that a user is more liberal, while a more positive score indicates that a user is more conservative. 
For this computation, we use a set of elites, including members of Congress, prominent members of the executive branch (e.g., the President, the Vice President, and various cabinet members), and prominent political organizations (e.g., the RNC and DNC Twitter accounts).
We collected the followers lists for each of these elites' accounts. 
To train the estimation model, we randomly select fifty thousand Twitter users who follow ten or more of the elites. 
We fit the correspondence analysis model on the correspondence matrix of the fifty thousand users and the elites. Each cell in the matrix indicates one if the user follows that elite and zero otherwise, specifying three component dimensions for the model output.
We use the fitted correspondence analysis to project scores into the same ideological space for the remaining users who follow at least three members of the model's elites. 
We standardize these scores such that the average is zero, and a score of one is one standard deviation of ideology away from zero. 

Finally, we assign partisanship to users based on their ideology scores. 
By connecting the linked panel to the user-generated ideology scores, we identify the deciles for each partisan group. 
We find that 90\% of Democrats have scores less than -0.35, and 90\% of Republicans in the linked panel are to the right of 0.04.
Thus, we classify the friends of the linked panel as Democrats if their estimated ideology is -0.35 or less and the user as Republican if their estimated ideology is 0.04 or greater. 
Users in between are considered non-partisans.
Using these classifications, we construct a measure similar to our offline isolation measures for users' partisan isolation in their online networks:

\[
\text{Online isolation} = \frac{\sum_{k=1}^{n}\mathbb{P}\mathbb(p_k=p_i)}{\sum_{k=1}^{n})}
\]
where $k$ is a given friend of the matched user $i$\, $p_k$ is the partisanship of the friend, $p_i$ is the partisanship of the matched user as estimated using the method described above,  and $n$ is the number of Twitter friends from user $i$ for whom we were able to estimate a ideological score.  We use this quantity as our measure of online partisan segregation in the following analyses. We limit the analysis to Twitter-L2 matched users for whom we can calculate the ideology for at least 10 of their friends 
\footnote{In the supplemental materials (Section 8), we show that this decision does not affect the median levels of online isolation. The main effect of using this cut-off is to avoid users whose networks are not informative.}.  

To summarize, both our measures serve as proxies for the number of voters or social media users in the matched user's offline and online networks who are co-partisans. In the following analysis, we assume that these measures exist on the same scale, allowing us to compare the levels of partisan segregation for the same voter both online and offline. 
For example, if a user's offline isolation is 0.33 and their online isolation is 0.33, we would assume that these lead to equivalent amounts of exposure to and interaction with out-partisans online and offline.

\subsection*{Ethical Considerations for Data Collection}
The linkage of Twitter data with voter file data raises privacy concerns for the users contained in the data, particularly as users (and voters) may not expect their data to be linked in this way. 
As a first step, we only link individuals in the voter file to publicly available Twitter accounts, meaning private Twitter accounts are not included in the data.
Second, given the voter file contains fine-grained demographic and residential information, we take seriously considerations of data security and privacy for this data to protect the personally identifiable information (PII) of the individuals contained in the data. 
Our data management plan was approved by New York University's IRB FY2021-4869. 
PII is stored on secure servers with strict access control only allowing read access to individuals on the project.
In this manuscript and other outputs of this project, we aggregate the data such that no particular individual can be re-identified from these aggregated statistics.

\section*{Results}
We first show the levels of offline and online isolation for the population of matched Twitter users, and the difference between Democrats and Republicans. 
We next analyze the differences across voters living outside metropolitan areas, in minor metropolitan areas, and in major metropolitan areas across a wide set of socio-demographic subgroups. Lastly, we discuss the correlation between the online and offline measures of ingroup isolation. 

\subsection*{Population-level offline and online isolation}

\begin{figure}[!htpb]
  \captionsetup{justification=raggedright, singlelinecheck=false}
  \caption*{\large\textbf{A}}
\scalebox{.9}{\begin{tabular}{c}
\includegraphics[width=1\linewidth]{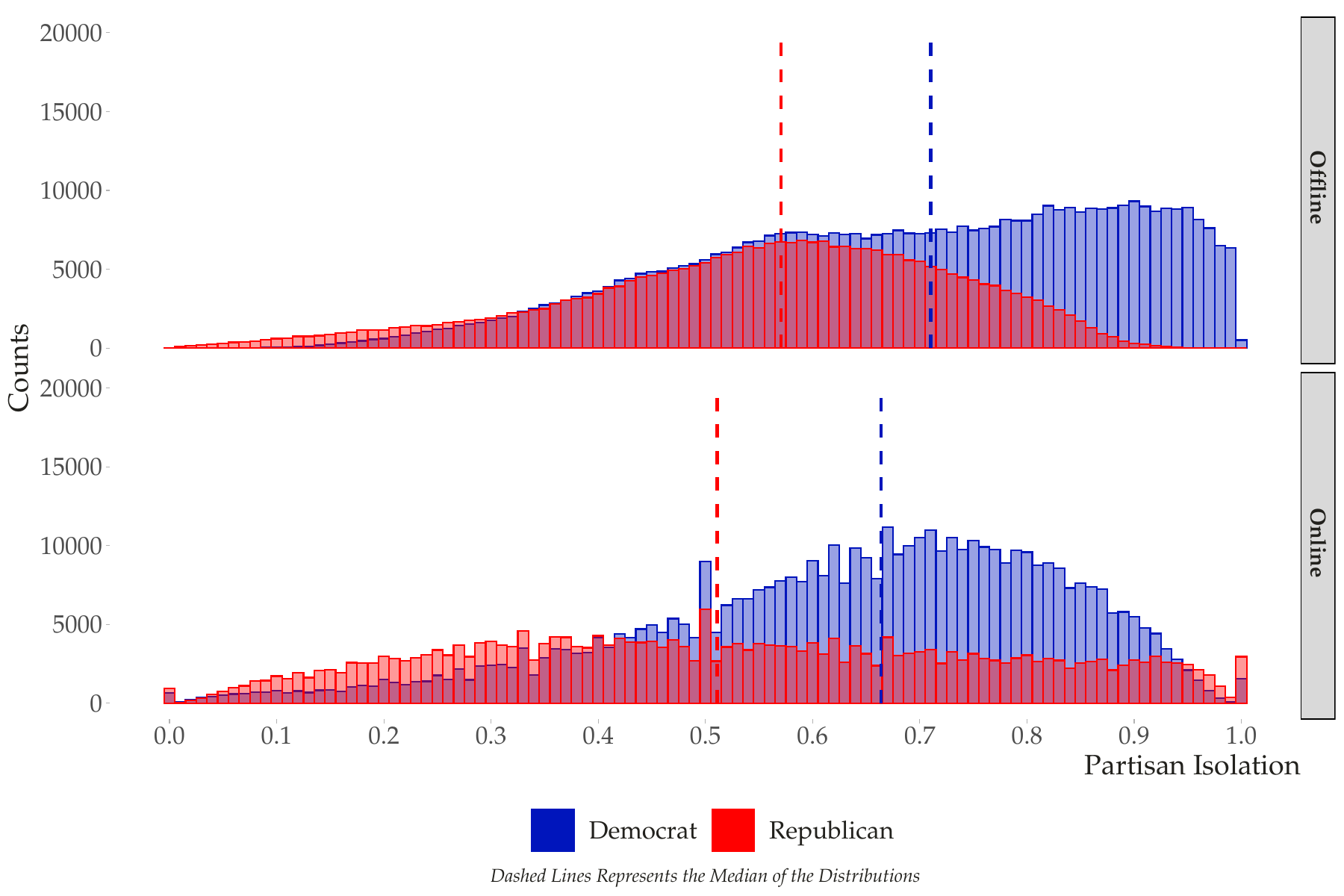} \\
\end{tabular}}
  \captionsetup{justification=raggedright, singlelinecheck=false}
  \caption*{\large\textbf{B}}
\scalebox{.9}{\begin{tabular}{c}
\includegraphics[width=1\linewidth]{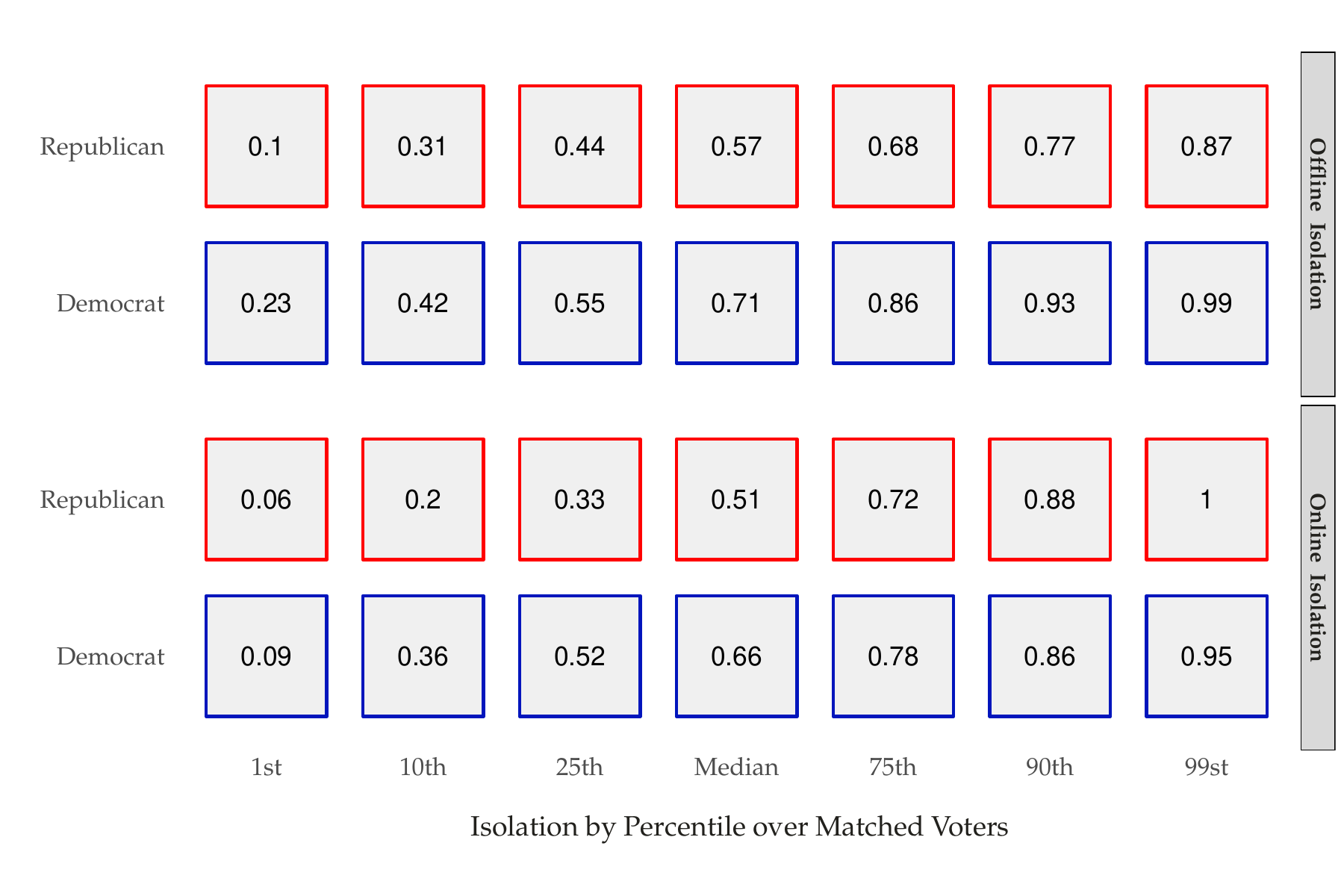} 
\end{tabular}}
\caption{\textbf{Partisan Offline  and Online Isolation}. Democrats are shown in blue, and Republicans are in red. Plot A presents the full distribution of values for isolation to ingroup split between Democrats and Republicans. Plot B shows the same distributions across percentiles for one million linked voters.}
\label{fig:hist}
\end{figure}

We first describe the distribution of online and offline isolation by partisanship. 
In Figure~\ref{fig:hist}, Plot A shows distributions of offline  (top) and online (bottom) ingroup isolation for Democrats (blue) and Republicans (red). 
We plot the median (dashed line)  for both Republicans and Democrats. 
Plot B shows the values at the 1st, 10th, 25th, 75th, 90th, and 99th percentiles. These quantities allow us to understand at distinct points of the distribution the differences between online and offline environment for Democrats and Republicans
Within each sub-plot, the top facet corresponds to offline isolation, the bottom row corresponds to online isolation, red represents Republicans, and blue represents Democrats. 

We find that Democrats have higher average levels of segregation (more isolated within ingroup communities) than Republicans, both online and offline.
And we find that for both groups, the median levels of isolation are greater for the offline distributions than the online distributions. 
For Democrats, we find that the median offline partisan isolation is 0.71, and the median online partisan isolation is 0.66.
For Republicans, we find a median isolation of 0.57 for offline isolation  and 0.51 for online isolation. 
 
These numbers also show that Democrats are more isolated offline than are Republicans (note the relative median values of 0.71 versus 0.57), consistent with previous research \citep{brown2021measurement}. And these differences also exist in their online environment, where the relative median values are 0.66 for Democrats compared to 0.51 for Republicans. This indicates that the median Republican Twitter user has in  online environment almost equaly divided between in-partisans and out-partisans (0.51 in ingroup isolation), which is quite different than the traditional notion of an echo chamber.

% code/descriptive_analysis_imputation_no_weights.r
%%% remove ingroup from the facets and labels. 

Beyond median differences in segregation, Plot B of Figure \ref{fig:hist} highlights important dynamics at the extremes of the distributions. At the 90$^{th}$ and 990$^{th}$ percentiles, online isolation approaches offline levels, and partisan gaps narrow. Both Republicans and Democrats show that nearly 90\% of their online friends are co-partisans at these extremes. Notably, at the 90$^{th}$ percentile, Republicans experience higher isolation online than offline. These cases illustrate that, at the extremes, online environments {\it can} resemble echo chambers, with over 80\% of one’s network composed of ingroup members. We note that the nature of the online environment is a choice by the individual, and we might expect that some number of people would choose homogeneous environments.

At the lower end of the distributions, however, patterns diverge. Republicans tend to experience lower online isolation than Democrats, with greater exposure to outgroups. By the 25th percentile, Democrats already show majority-ingroup networks (isolation > 0.5), while Republicans at the same percentile remain exposed to liberal users, with about two-thirds of their online ties connecting to Democrats.

\subsection*{Correlates between Online and Offline Isolation}
\label{sec:indiv}

\begin{figure}[!htpb]
\begin{tabular}{cc}
\includegraphics[width=.5\linewidth]{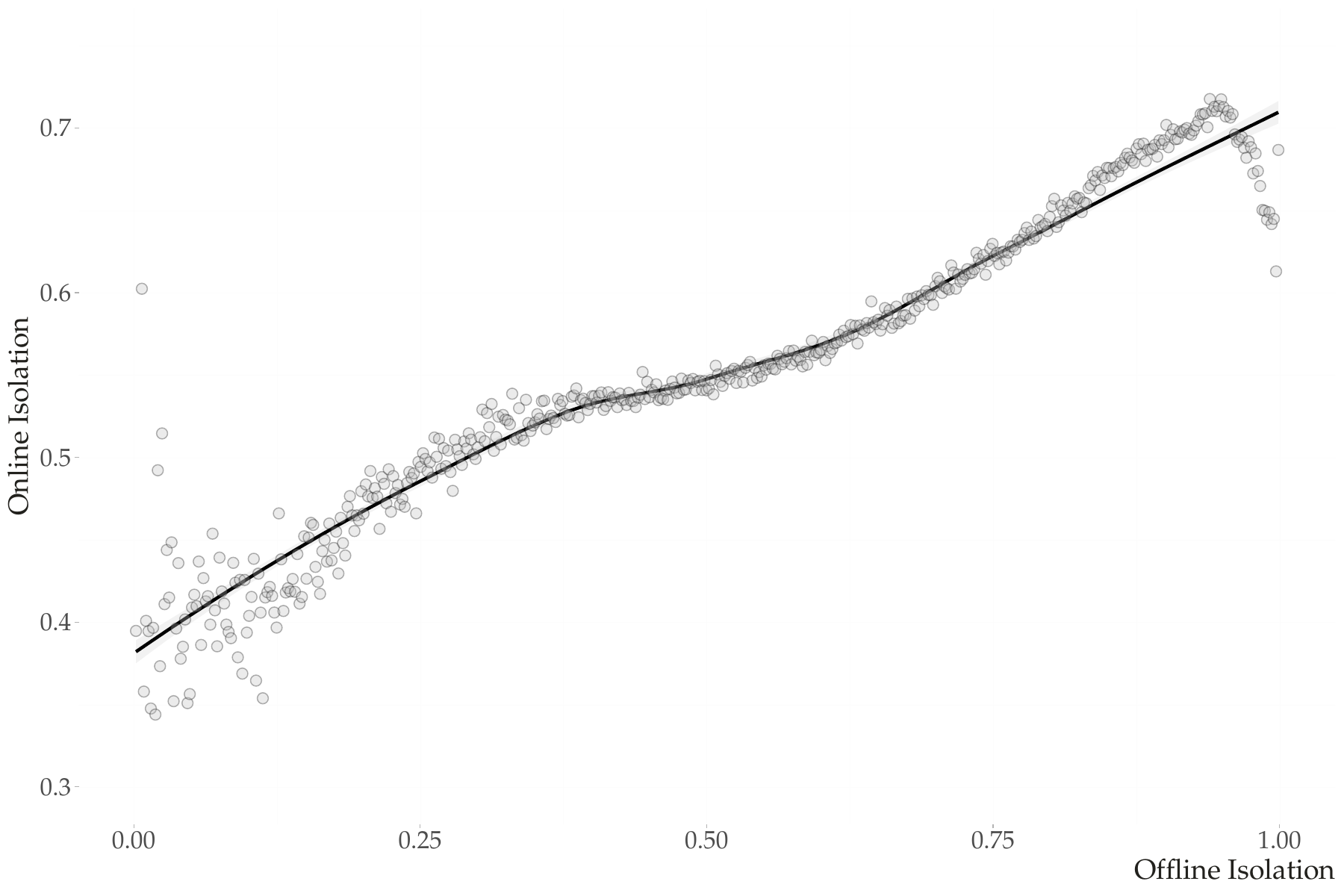}  &
\includegraphics[width=.5\linewidth]{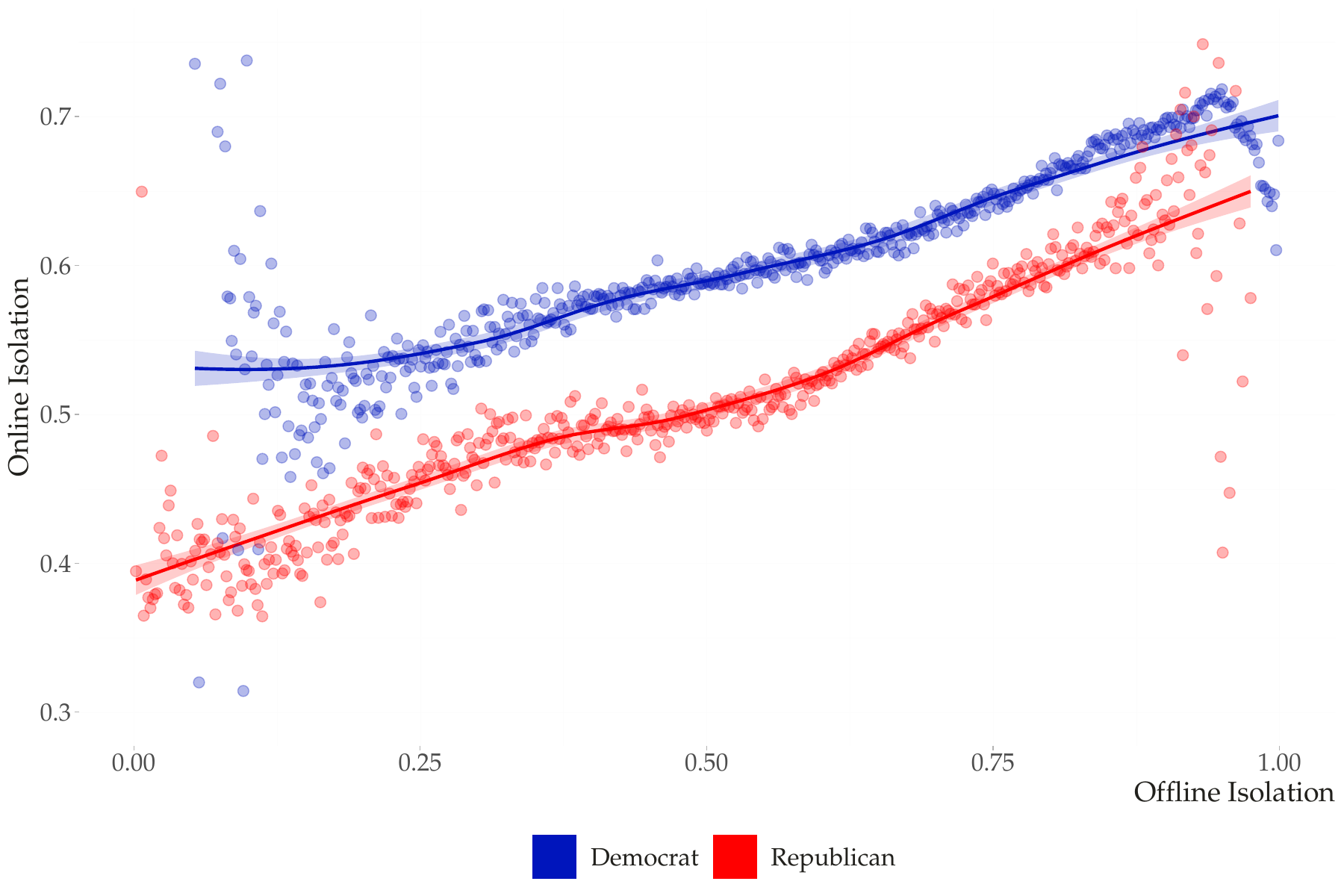} \\
a) All & b) Democrats vs Republicans
\end{tabular}
\caption{\textbf{Relationship between Online and Offline Isolation}. Democrats are shown in blue, and Republicans are in red. Each dot represents average online isolation over a hundred binned values of offline Isolation. Lines represent a smoothed non-linear function over the points}
\label{fig:corr_off_on}
\end{figure}

We now examine the direct relationship between offline partisan isolation and online partisan isolation. 
In Figure~\ref{fig:corr_off_on}, we show the relationship between offline (x-axis) and online isolation (y-axis). 
We bin offline  isolation values at five hundred equal intervals from the variable distribution, compute the average for offline isolation for each bin, and plot the resulting values. 
Lines on the figure represent a smoothed non-linear function over the averaged points. 
On the left panel, we plot offline (x-axis) and online isolation (y-axis) for all users.
On the right panel, we plot similar information but for Republicans (red) and Democrats (blue) separately.

For all users, regardless of partisanship, we find that as offline isolation increases, changes in the online isolation to ingroup social media users follow accordingly. That is, users with lower offline partisan isolations tend to have lower levels of online partisan isolation. In other words, users with more politically diverse real world neighbors have a  more diverse set of online friends (i.e., users followed on Twitter), with a higher presence of outgroup users in their Twitter networks.  And we note that this is not an artifact of cross-partisan differences.  When we examine Democrats and Republicans separately (panel (b)), we see the same phenomena (though the relationship is not as strong).

% code/connect_with_census

We further explore differences in online and offline partisan segregation across demographic and political subgroups using the rich set of variables collected in the L2 voter files. 
Because offline segregation varies by race, party, population density, and other demographics, these analyses allow us to see  the degree to which online sorting follows offline patterns, and also how the population-wide differences depicted earlier hold for specific socio-demographic groups.

\begin{figure}[!htpb]
\scalebox{1}{\begin{tabular}{c}
\includegraphics[width=1\linewidth]{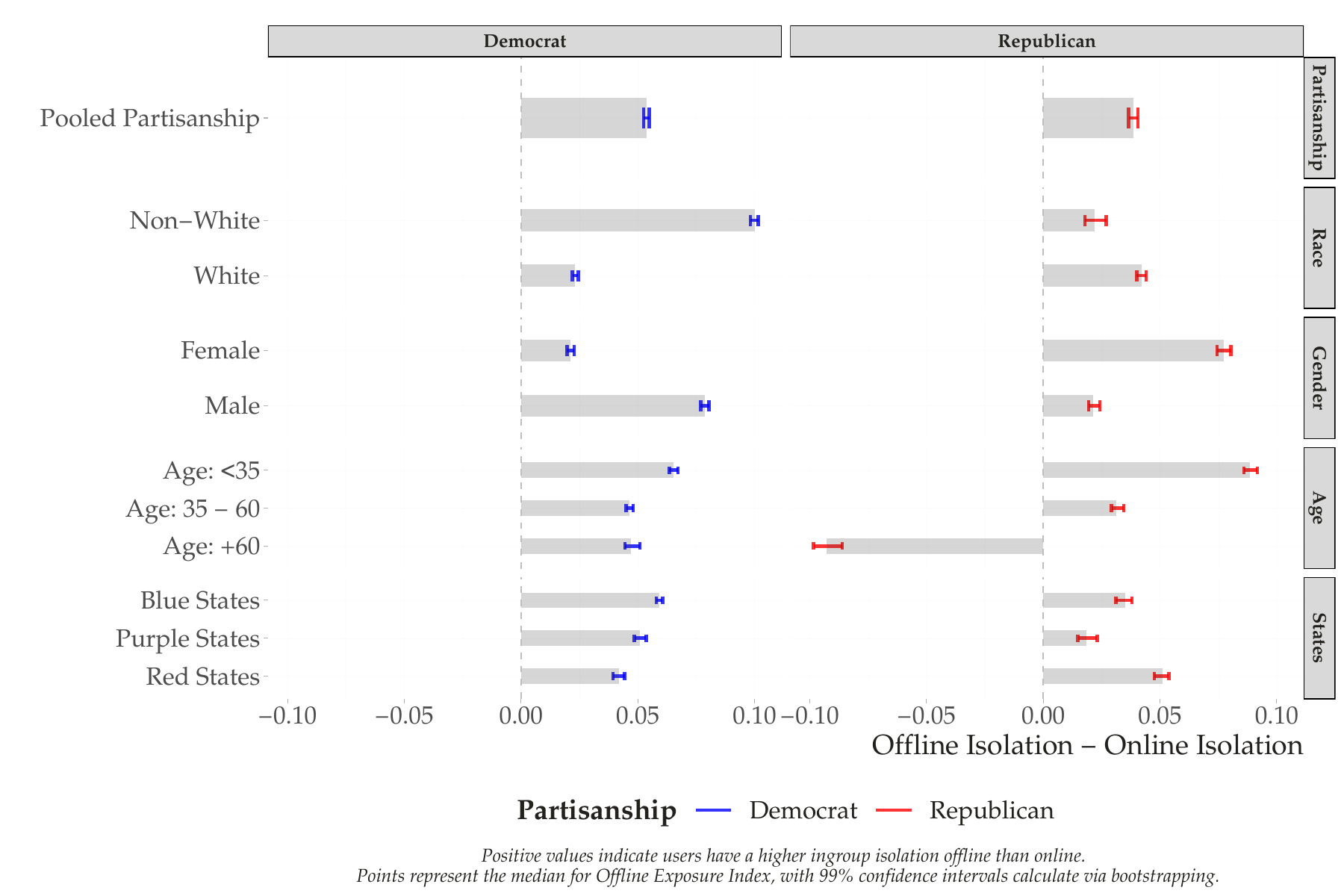}
\end{tabular}}
\caption{\textbf{Partisan Offline and Online Isolation Across Socio-Demographic Subgroups}. Point estimates represent the median value of the difference between online and offline ingroup isolation for each group with 99\% confidence intervals calculated via bootstrapping. Positive values indicate voters who have higher offline isolation than online isolation. Democrats are shown in blue, and Republicans are in red.}
\label{fig:subgroup}
\end{figure}

%% Demographics
We show how these differences vary across distinct demographic groups. In Figure \ref{fig:subgroup}, we show median paired distances between individuals' offline and online isolation across various sociodemographics and partisanship. 
Positive values indicate users experience higher levels of offline segregation than online segregation.  
Across almost all subgroups, including gender, race, age, partisanship, and state type (Republican-leaning states, Democratic-leaning states, and swing states) \footnote{Swing states are defined as states that have a partisan vote-share split between Biden and Trump at 3\% or less, meaning that if in 2020, a state's vote share for Biden compared to its vote share for Trump was within three percentage points of each other, we define that state as a swing state. Using this definition, Arizona, Florida, Georgia, Michigan, Nevada, North Carolina, Pennsylvania, and Wisconsin are classified as swing states.} users experience considerably higher levels of isolation offline than online. This supports our overall argument that local echo chambers are more pronounced than online experiences are robust across most subgroups. 
The main exception to this pattern is Republican voters who are older than 60 years.
For this sociodemographic group, their levels of online segregation are higher than their offline isolation. 
This is the only group in which their online experience can actually be identified as living in more of an echo chamber online than offline. 
This finding is interesting considering the consistent evidence that older Republicans are often higher consumers and spreaders of low-quality information on social media \citep{guess2019less,grinberg2019fake,guess2021cracking, gonzalez2023asymmetric}.

% code/generate_graphs_differences_subgroups.r

In the SM Section 2,  we present additional analysis exploring differences in online and offline partisan segregation across distinct levels of urbanization and population density. 
Differences between online and offline isolation hold among most areas in the United States, regardless of the population density and urbanization. The exception resides in the extremes: in high-density areas, Republicans have more ingroups in their online than offline networks, and in very low-density areas, Democrats show higher online isolation. In summary, in the extremes of high- and low-population-density areas, social media enables both Democrats and Republicans to connect with other ingroup voters, who are not as easily accessible within their local bubbles. 

Furthermore, as discussed in the methods section, our empirical analysis involves several measurement choices that may affect our results. 
These choices involve the values for the nearest neighbors calculation, the use of ideology scores based on Twitter users' followers' networks, and the cut-off values to classify users as Democrats and Republicans. 
To validate our results,  we provide several sensitivity analyses showing the robustness of the main findings drawn from our study across distinct measurement choices. 
In the supplemental materials, we demonstrate that the results remain unchanged when we consider only the 500 nearest neighbors when calculating measures of partisan offline isolation. 
We also present in the supplemental materials results using a discrete assignment of partisanship, where the partisan identity is assigned based on the highest probability from the imputation calculation, instead of calculating offline isolation based on the posterior probability for each partisan group. 
The results remain substantively unchanged, with Republicans being slightly more isolated using this alternative measure.

\section*{Discussion}

Social media is often seen as a primary driver of the formation of overly homogenous online communities - so-called echo chambers - in which citizens more easily isolate themselves from individuals and opinions from opposing parties and, consequently, reinforce their system of beliefs \citep{sunstein2018republic, bakshy2015facebook, flaxman2016}. 
While a larger body of research has provided a more nuanced view of the prevalence of the phenomenon \citep{eady2019many, guess2019less, gentzkow2011ideological, barbera2015birds}, these studies do not account for partisan segregation in the offline environment. Using a novel dataset connecting online and offline information for over a million Twitter users in 2021,  we deal directly with this concern by explaining how individuals' online and offline environments compare.

Our study presents four core findings. 
First, we find that for both Republicans and Democrats, partisan segregation is {\it higher} in their offline environments than in their online environments. 
That is, even though their online environments are homogeneous with more politically like-minded people than not, their offline environments are \textit{even more} politically homogeneous. In other words, contrary to popular concerns about echo chambers, social media can actually increase exposure to the outgroup online relative to users' online environments. 
Second, we found that in both offline and online environments, segregation is higher for Democrats than Republicans, meaning that in both online and offline environments, Republicans are more frequently exposed to outpartisans than Democrats.
Third, this finding holds for most socio-demographic subgroups. The only subgroup for which this does not hold is older Republicans who are more isolated among copartisans online than in their offline environments. 
Lastly, online partisan segregation is positively correlated with offline partisan segregation.  

Our results make contributions to three distinct areas of scholarly inquiry. First, our findings speak to studies on online echo chambers and social media homophily \citep{barbera2015birds, eady2019many, bakshy2015facebook, tornberg2022digital}. Prior research in political communication has debated the extent of ideological “bubbles” in social networks, with evidence that partisan homophily drives tie formation online. We contribute to this literature with a direct comparison of online and offline sorting. While we do find that social media users curate politically homogeneous communities (consistent with earlier studies), we are able to add the important observation that the degree of ideological isolation online is  \textit{lower} than in real-world geographical spaces. Only in the upper part of the distributions do voters experience more of an echo-chamber online than offline. For the most part, social media exposes users to more out-group users than do their offline interactions.\footnote{We note that our offline network measure does not directly measure interactions, but our survey analysis (Figure~\ref{fig:corr}) suggests it is capturing this.}

Second, our results contribute to the emerging literature on integrating offline context with online political behavior, providing novel evidence about partisan sorting both online and offline \citep{gentzkow2011ideological, barbera2015birds, eady2019many, bakshy2015facebook}. Except for \cite{gentzkow2011ideological}, which uses survey data to measure offline partisan segregation, most studies of online partisan sorting rely solely on online information to describe the prevalence of online echo chambers and the role of homophily on online network composition.  To our knowledge, ours is the first study to bring the dynamic of online segregation to this discussion, engaging with a robust scholarship on spatial partisan sorting in American politics \citep{brown2021measurement, enos2017space, mason2018uncivil, bishop2009big}, and propose a baseline to which we could empirically assess levels of online sorting among the American electorate.\footnote{One other excpeption is \cite{asimovic2021testing}, which although it was not focused on measuring online and offline segregation, did actively compare the two in Bosnia as part of an effort to explore the heterogeneous treatment effects of Facebook usage on out-group ethnic attitudes in a deactivation experiment.}

Lastly, our article also speaks to studies focusing on the centrality of partisan geographical segregation in the United States \citep{bishop2009big, levendusky2009partisan, enos2017space}.
In a recent article, mobilizing an empirical toolkit similar to the data described in our paper, \citet{brown2021measurement} present evidence of extensive partisan segregation in the US -- their results indicate that a large proportion of the American electorate lives with virtually no exposure to outgroup voters in their residential environment. 
This dynamic has been shown to affect mass and elite polarization \citep{bonica2014mapping}, cooperation for shared benefit across groups \citep{enos2016intergroup},  levels of trust in government and anti-system attitudes \citep{cramer2016politics}, and even adherence to scientific health recommendations during the COVID-19 pandemic \citep{baxter2022local}. 
By linking millions of voters’ residential contexts with their online networks, our study demonstrates that partisan segregation is considerably greater offline than online, indicating how offline partisan sorting is still a decisive feature of the american politics even in the years of social media. And, by showing the correlation between offline and online partisan isolation, we provide suggestive evidence that such offline segregation leads to online partisan segregation.

Our study is, of course, not without limitations. 
First, we assume that the measure of offline segregation is predictive of the extent to which an individual is actually exposed to the ingroup and outgroup offline. 
While we do validate this measure using a survey measure, this measure may miss important avenues of exposure to partisan information. 
For example, this measure does not include interactions in the offline environment, such as school board meetings, yard signs, and community projects, social and religious groups, among others, that could lead to exposure to other partisans or partisan information. 
Second, we assume that the measures we employ are roughly equivalent. 
That is, if an individual is measured to have a 0.33 level of ingroup isolation offline and 0.33 online, then these would lead to proportionally the same amount of exposure to in-partisans and out-partisans. 
But data limitations in both the offline and online world prohibit us from meaningfully measuring the equivalent values. 
Despite these limitations, we believe the proxies for the measures that we employ provide valuable insights into the dynamics of offline and online partisan segregation, and this is especially the case when making comparisons within offline and online isolation analyses. 

%% Megan see this pararagraph. 
Lastly, our study provides a snapshot of the differences between online and offline partisan segregation using a specific platform, Twitter, at a particular moment in time. 
This consideration is particularly relevant given the significant changes Twitter has undergone in recent years following Elon Musk's acquisition \citep{bisbee2025vibes, barrie2022did}. 
Yet we do not believe this caveat diminishes in any way the importance of our findings.
First, our results are temporally valid to a point in time when Twitter functioned as a critical medium for the American public to learn and discuss politics, and as a platform on which politicians and the media relied heavily to communicate and learn about public preferences \citep{barbera2019leads, huszar2022algorithmic, flamino2023political}. 
Second, the vast majority of scholarly work on social media and politics produced in the last decade relies on Twitter data \citep{tucker2018social}, including studies that make claims about the role of social media in facilitating the formation of echo chambers.
Therefore, while it is true that Twitter is no longer the same as it was in 2021, our findings still provide a proper understanding of the social media environment, using a platform that was once critical for political communication. 

Our findings complicate the dominant narrative of social media as the primary determinant of whether people end up in ``echo chambers.'' 
While online spaces are often portrayed as uniquely isolating, our results show that partisan segregation is actually higher in Americans’ offline environments than in their online ones. 
For most users, social media provides more, not less, exposure to outpartisan perspectives relative to their local communities. 
At the same time, we uncover meaningful asymmetries: Democrats remain more isolated than Republicans across both settings, and older Republicans are the only group whose online networks are less diverse than their offline ones. 
Importantly, the positive correlation between online and offline isolation underscores that the digital and physical worlds are not independent spheres but intertwined environments, with online experiences often reflecting offline realities.

Taken together, these results suggest that the effects of social media on political attitudes and behaviors cannot be understood in isolation from the broader social and geographic contexts in which people live. 
Rather than treating social media as a standalone driver of polarization, our study highlights the importance of situating digital technologies within the context of offline partisan segregation. 
This study provides a more nuanced understanding of partisan segregation in the United States and highlights the need for future research to jointly consider online and offline contexts when assessing the political consequences of social media.

\newpage
\bibliography{bib}

\newpage
\setcounter{figure}{0}    

%\begin{document} 
	
	\begin{center}
	{\LARGE \textbf{The relationship between offline partisan geographical segregation and online partisan segregation.}}
	
	Supporting Information Files (SIF)
	
	\end{center}

%\tableofcontents

%\newpage
    \setcounter{page}{1} 
    
    \setstretch{1.64}
    \justify
    \setlength {\parindent}{15pt}

\section{Survey Materials}

These data are used to validate our measure of offline partisan segregation, which captures the extent to which an indivdiual's offline network is more ideologically diverse or segregated. 
Given that the measure we use in the main manuscript is based on an individual's nearest \emph{geographic} neighbors, it may not represent who they interact with. 
More specifically, it could be the case that participants live in a geographically diverse area but only spend time with out-partisans.
Thus, we ask the following questions of individuals to validate that their neighbors are a reasonable proxy for measuring their offline networks. 

\begin{figure}[!t]
\begin{tabular}{cc}
\includegraphics[width=.8\linewidth]{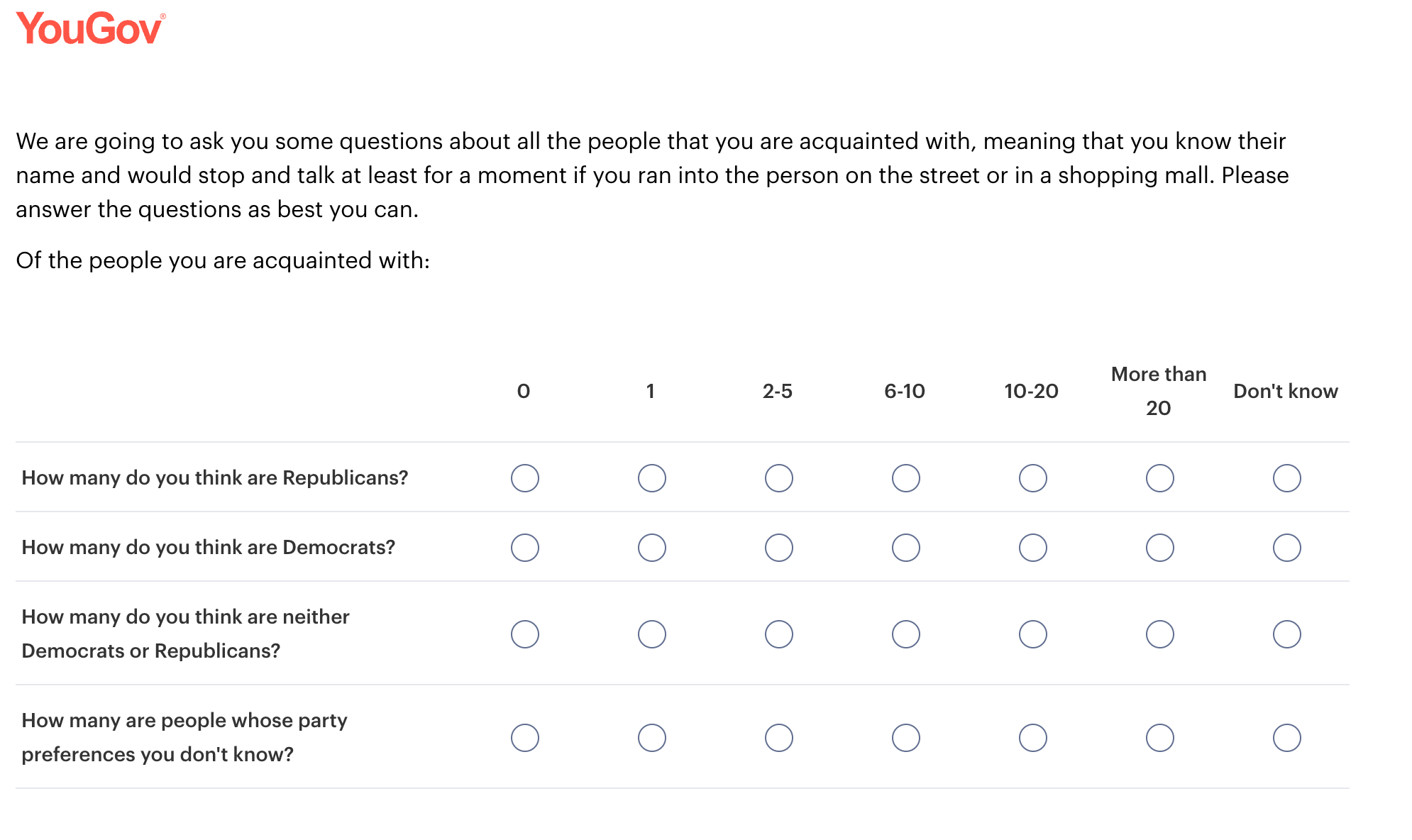}  \\
\end{tabular}
\caption{\textbf{Survey question for the ideological breakdown of an individual's acquaintences }}
\label{fig:surveyq_acquaintences}
\end{figure}

\begin{figure}[!t]
\begin{tabular}{cc}
\includegraphics[width=.8\linewidth]{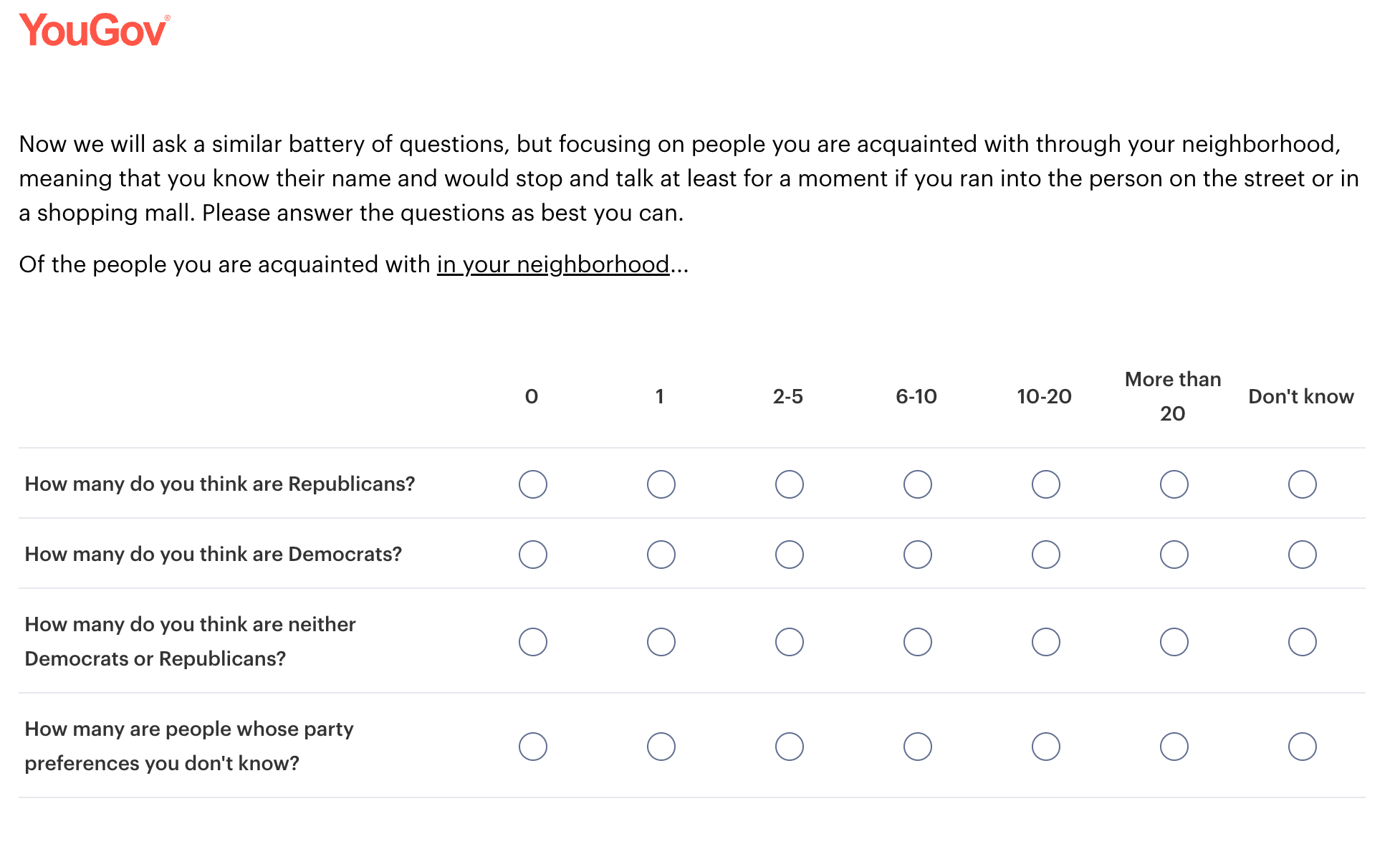}  \\
\end{tabular}
\caption{\textbf{Survey question for the ideological breakdown of an individual's neighbors }}
\label{fig:surveyq_neighbors}
\end{figure}

\begin{figure}[!t]
\begin{tabular}{cc}
\includegraphics[width=.8\linewidth]{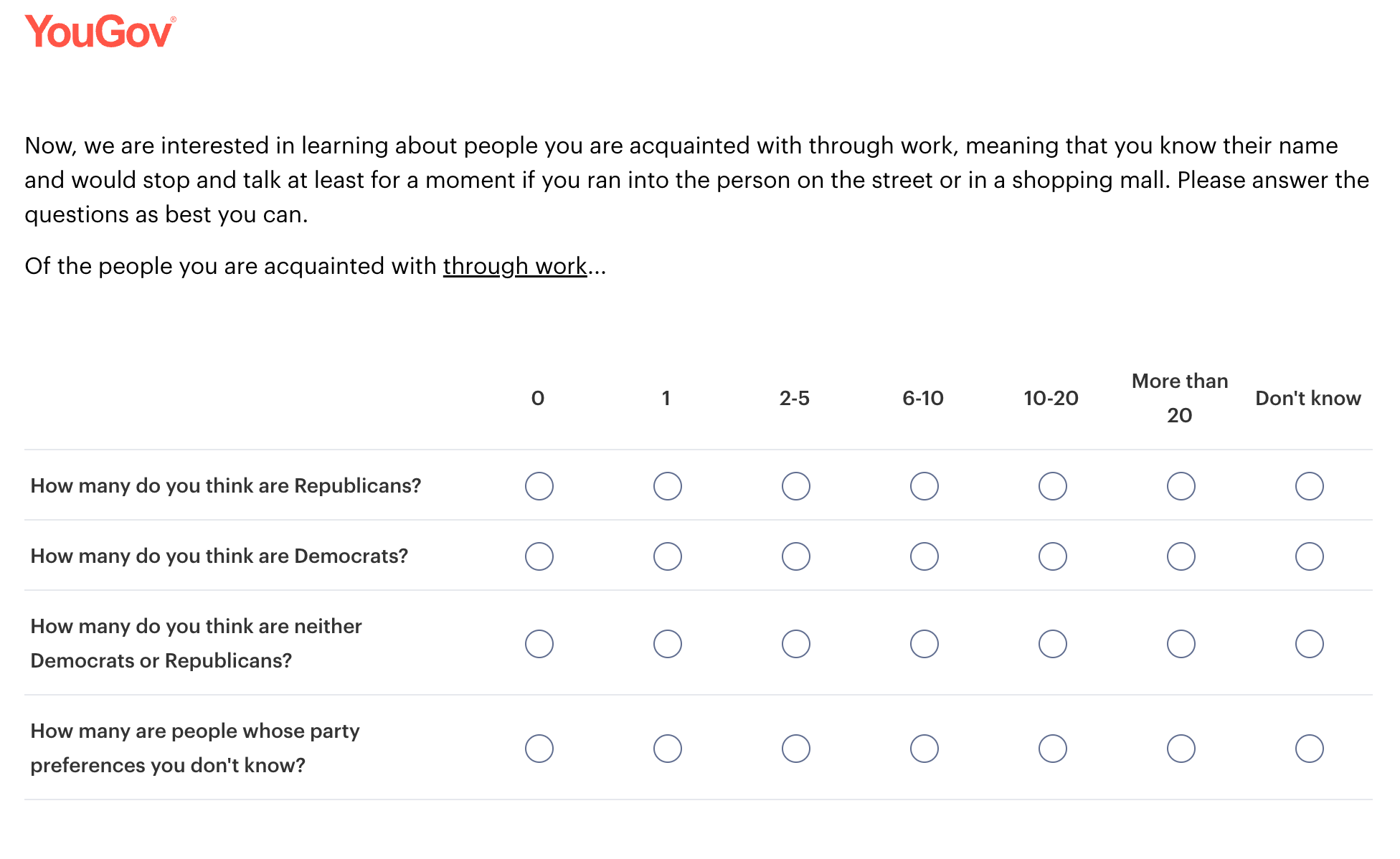}  \\
\end{tabular}
\caption{\textbf{Survey question for the ideological breakdown of an individual's coworkers }}
\label{fig:surveyq_coworkers}
\end{figure}

We fielded a survey of 3,500 respondents through the survey firm YouGov in early 2023. 
Respondents were recruited from adult US population, oversampling for social media users. 
For each respondent, we asked a battery of questions related to the ideological composition of their networks using a grid of questions.
These questions were adapted from \citep{gentzkow2011ideological}, which estimates the partisan makeup of a respondent's network. 
We include figures showing what the participants saw during the survey. 
In Figure~\ref{fig:surveyq_acquaintences}, we show the survey question related to the partisan makeup of an individual's acquaintances. 
In Figures~\ref{fig:surveyq_neighbors}~and~\ref{fig:surveyq_coworkers} we show the same for the partisan makeup of an individual's neighbors and the people they are acquainted with through work, respectively. 

\section{Comparing Online and Offline Isolation Across Urban and Rural Areas and Population Density}

\begin{figure}[!htpb]
\scalebox{.9}{\begin{tabular}{c}
\includegraphics[width=1\linewidth]{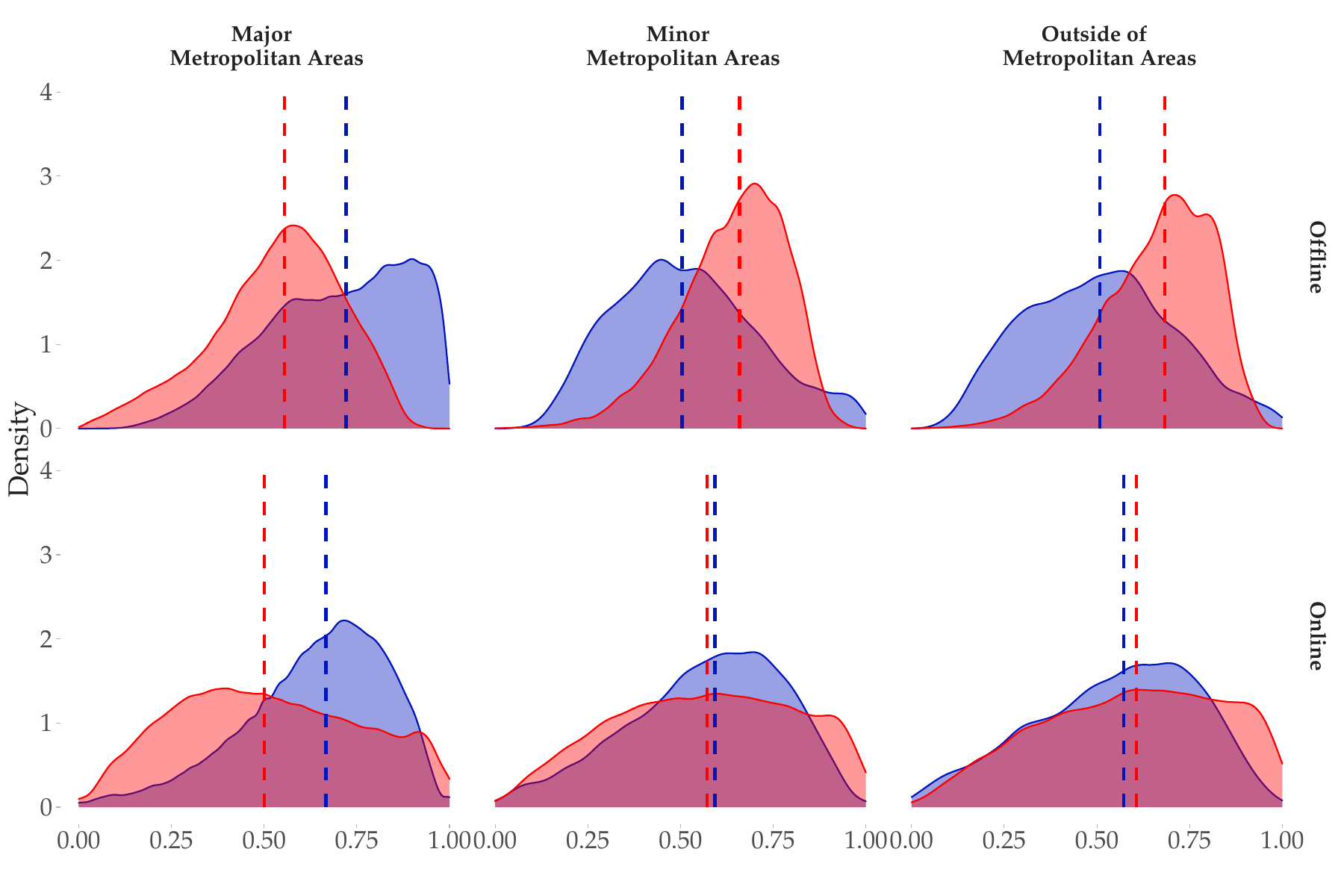} \\
\includegraphics[width=.95\linewidth]{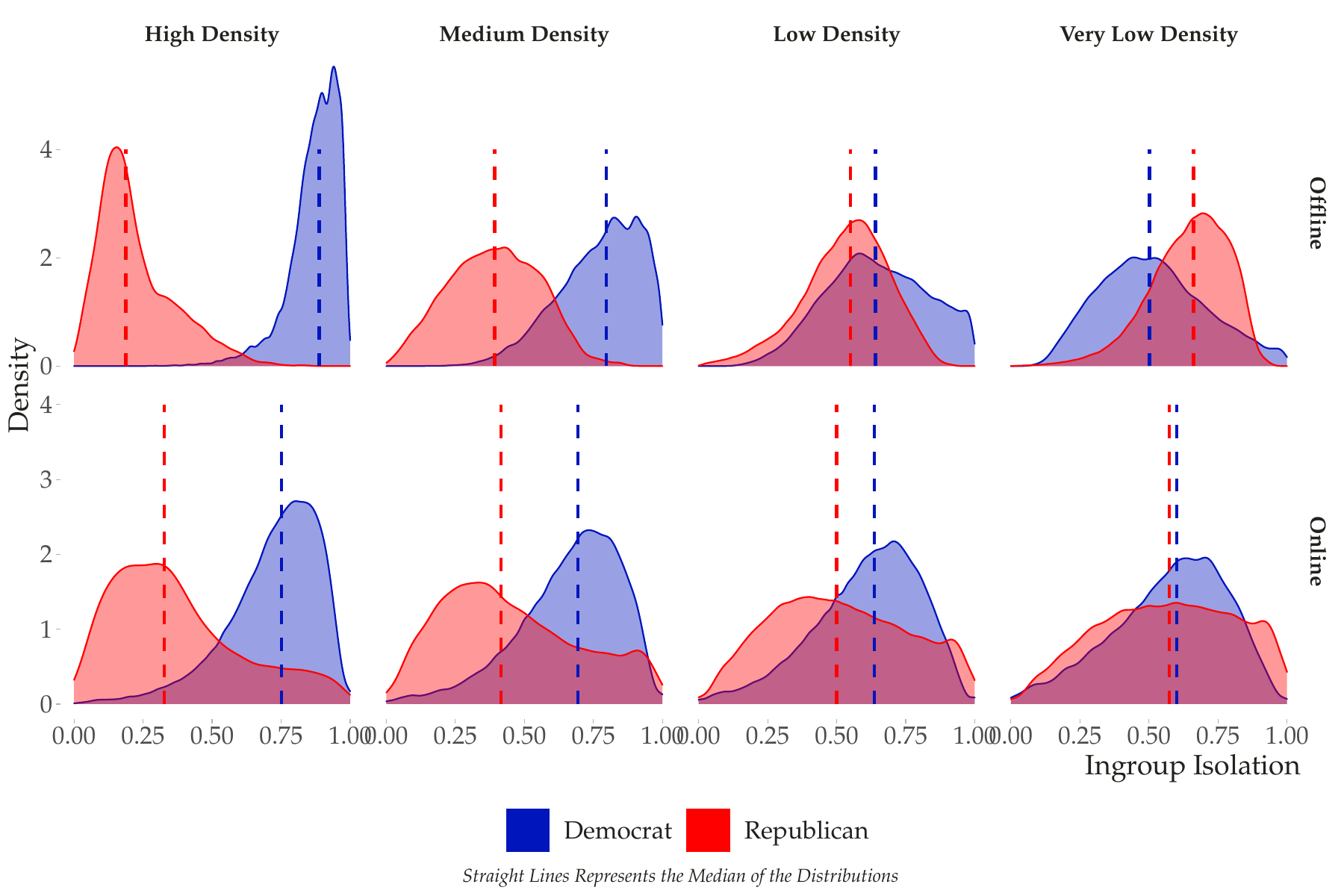} 
\end{tabular}}
\caption{\textbf{Partisan Offline and Online Isolation Across Metropolitan Areas and Population Density}. Democrats are shown in blue, and Republicans are in red. The upper plot presents the densities for isolation to ingroup split between Democrats and Republicans across Major, Minor, and Outside Metropolitan Areas using CBSAs. The bottom plot shows the distribution according to population density in each census tract}
\label{fig:metro}
\end{figure}

Figure \ref{fig:metro} presents differences in online and offline isolation across levels of urbanization using core-based statistical areas (CBSAs) to identify major metropolitan areas, and levels of population density \footnote{Measure of Population Density were adapted from D. H. Montgomery of CityLab. See measures here: \url{https://github.com/theatlantic/citylab-data/blob/master/citylab-congress/methodology.md}}
Democrats exhibit their highest levels of offline segregation in major metropolitian areas and tracts with high population denisity. 
Democrats who are in highly segregated offline enviornments also tend to be in more segregated online environments. 

As shown for the general voting population (not only our sample of twitter matched voters),  Democrats exhibit their highest/lowest levels of segregation in major metropolitan areas and tracts with high population density \citep{brown2021measurement}. We observe similar findings among our sample of matched voters, and more interestingly, these differences closely align with Democrats' online experiences. Democrats, who are highly segregated offline in major metropolitan areas, are also highly isolated on social media. In other words, their online experiences closely match their levels of segregation offline. For Republicans, a similar match appears, but on the other side of the distribution. For those Republicans living in high population density areas, their levels of offline isolation particularly small (less than 20\% of ingroup among their nearest neighbors), as well as their online segregation (around 25\% of ingroup in their Twitter networks).

A distinct pattern emerges when looking at differences between offline and online segregation outside of major metropolitan areas and more rural parts of the United States with low population density. In minor metropolitan and outside of metropolitan areas (mostly rural tracts) with a very low population density, the offline isolation of Republicans becomes larger than the offline isolation of Democrats. More interestingly, the correspondence between offline and online segregation changes in these rural areas. In these areas, Democrats show higher isolation levels online than offline -- a distinct partner from our population estimates. Social media for democrats living in highly Republican areas allows these users to break their ``local bubbles`` and give these voters means to communicate with others more aligned with their partisan identity. This is not the Republican experience. Even for voters living in these rural spaces, levels of isolation for republican voters are still smaller than their offline isolation. Even in these highly segregated areas, Republicans are far from experiencing online a echo-chambers that matches their offline lives.

%code/generata_graph_subgroups

\section{Comparing Online and Offline Isolation Across Subgroups}

In the main paper, we provide median differences and statistical tests to compare online and offline isolation across demographic groups (race, gender, and age). In this section, we present the reader with the full distribution for online and offline isolation across these subgroups. 

% add partisanship
% code/descriptive_analysis.r
\begin{figure}[!htbp]
\begin{tabular}{c}
\includegraphics[width=1\linewidth]{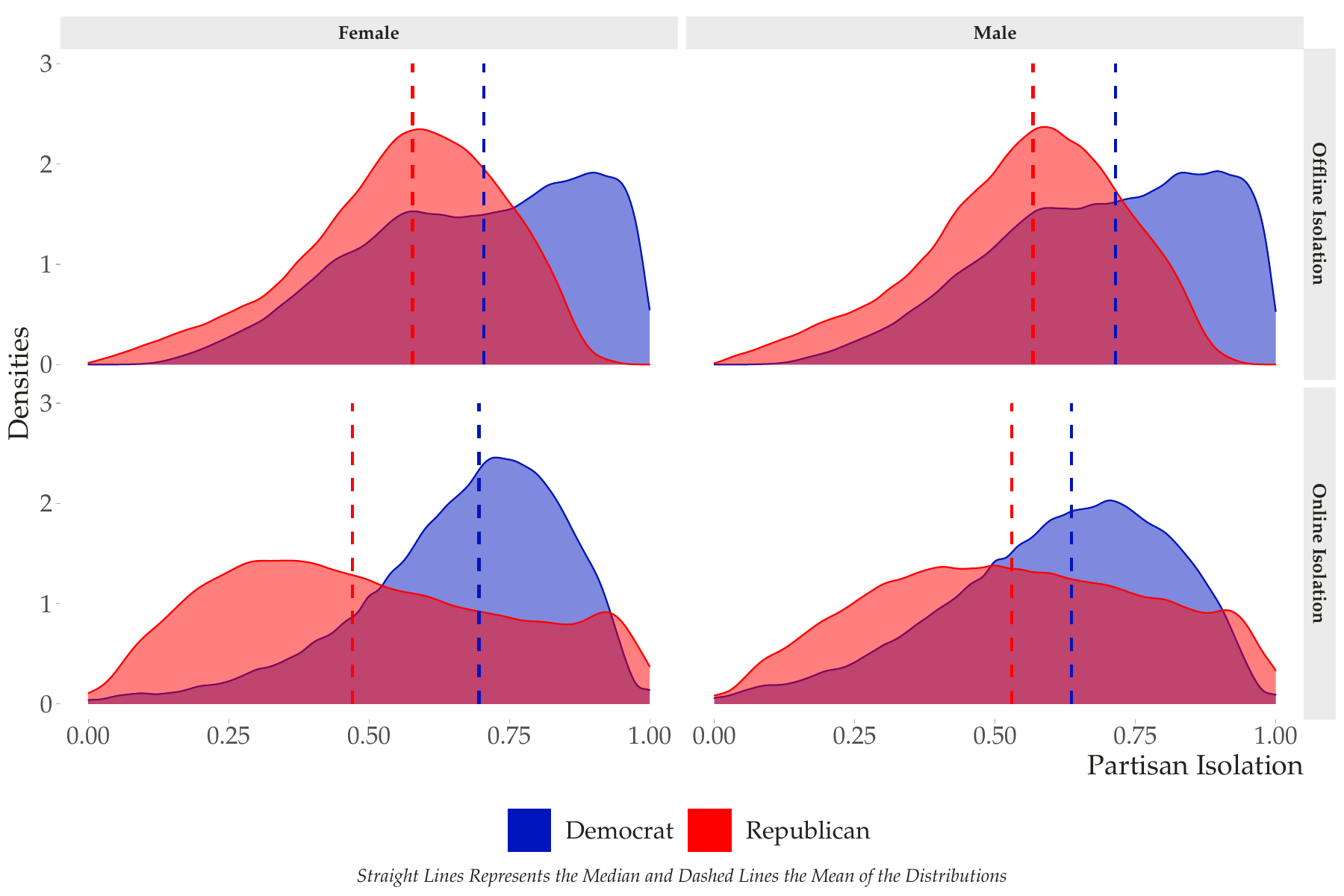}  \\
\end{tabular}
\caption{\textbf{Online and Offline Isolation by Voter's Gender}}
\label{fig:nope}
\end{figure}

%\newpage
% code/descriptive_analysis.r

\begin{figure}[!htbp]
\begin{tabular}{c}
\includegraphics[width=1\linewidth]{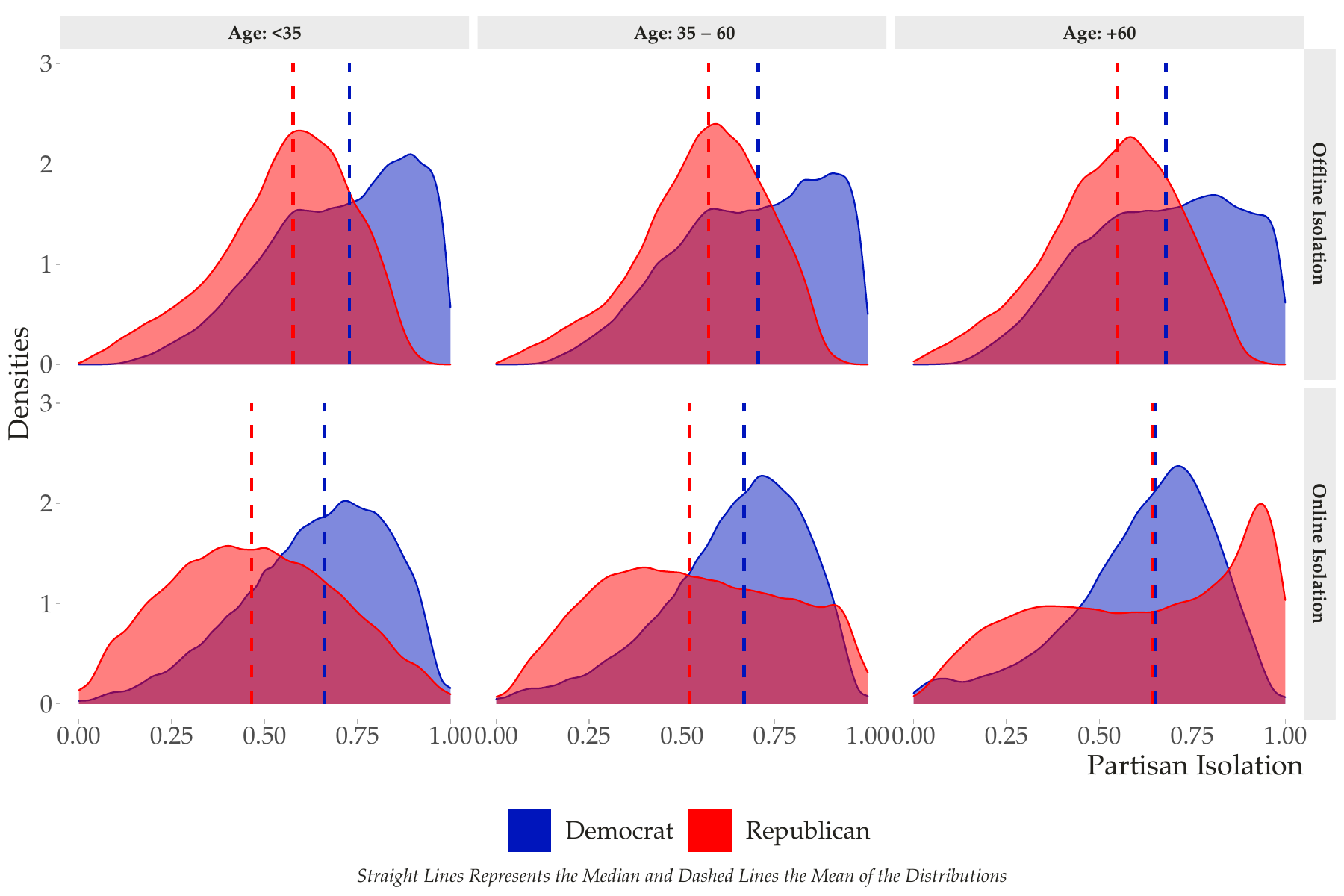}  \\
\end{tabular}
\caption{\textbf{Online and Offline Isolation by Voter's Age}}
\label{fig:nope}
\end{figure}

%\newpage
% code/descriptive_analysis.r

\begin{figure}[!htbp]
\begin{tabular}{c}
\includegraphics[width=1\linewidth]{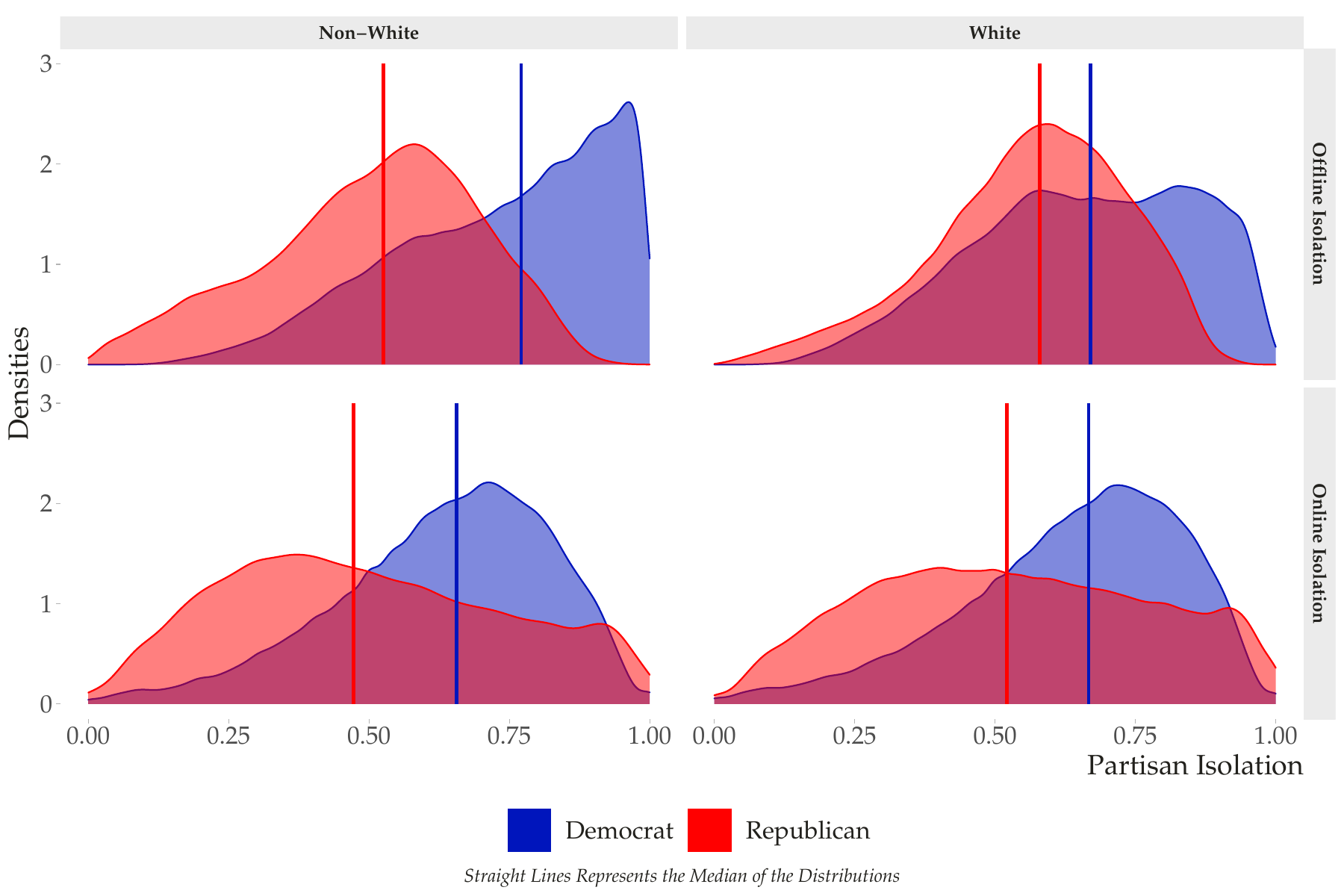}  \\
\end{tabular}
\caption{\textbf{Online and Offline Isolation by Voter's Race}}
\label{fig:nope}
\end{figure}

\section{Robustness Checks: Offline Isolation Under Different Cut-offs}

In this section, we replicate Figure 2 Plot B from the main paper using a distinct cut-off to calculate offline isolation. In the main article, we use the first 1,000 closest neighbors to estimate offline partisan segregation. Here, we present results using the 500 closest neighbors. Levels of offline isolation remain similar to those reported in the paper.

% code/descriptive_analysis_imputation_no_weights.r
\begin{figure}[!htpb]
  \captionsetup{justification=raggedright, singlelinecheck=false}
  \caption*{\large\textbf{Number of Nearest Neighbors = 1000}}
\scalebox{.9}{\begin{tabular}{c}
\includegraphics[width=1\linewidth]{figs_imp/probs/noweights_imp_offline_isolation_quantiles_circles.pdf} \\
\end{tabular}}
  \captionsetup{justification=raggedright, singlelinecheck=false}
  \caption*{\large\textbf{Number of Nearest Neighbors = 500}}
\scalebox{.9}{\begin{tabular}{c}
\includegraphics[width=1\linewidth]{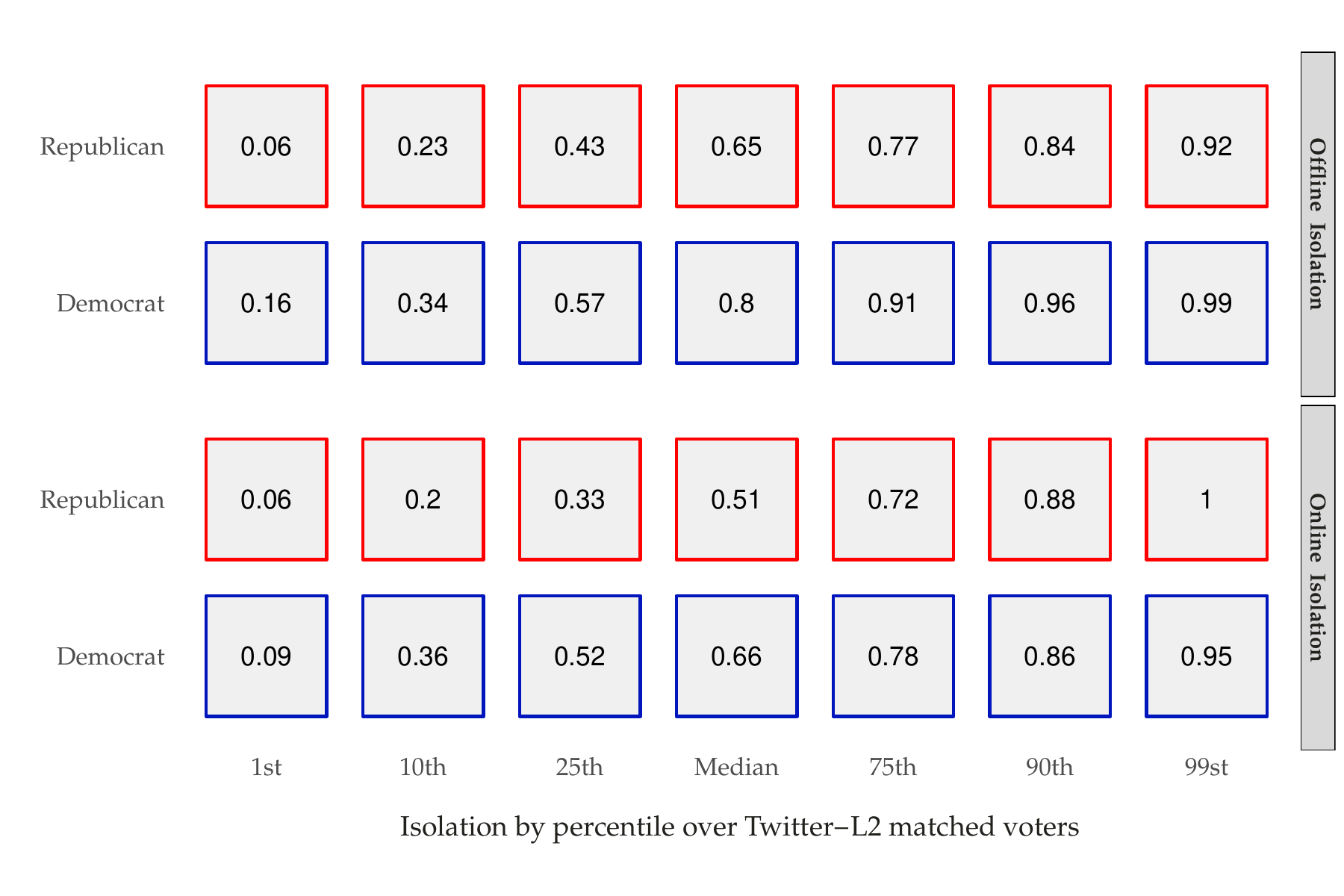} 
\end{tabular}}
\caption{\textbf{Partisan Offline  and Online Isolation}. Democrats are shown in blue, and Republicans are in red.}
\label{fig:hist2}
\end{figure}

\section{Imputation for Voters' Partisan Identity}

As discussed in the main manuscript, an important challenge of working with information from voter files, compared to survey data, for example,  is to deal with voters who are not registered as Democrats or Republicans. 

In this section, we describe our procedure and results to impute values for partisan identity for all voters from our augmented data set, as well as their neighbors.  Our imputation technique follows \cite{brown2021measurement} using a similar three-step procedure to impute partisanship for matched voters and their neighbors. 

In step one, we use partisanship information provided by the voter file vendor L2 to classify all voters. This classification comes from two distinct processes. On one side, thirty states ask voters to declare their party identification in the voter registration. Additionally, for states where party identification is not requested, L2 develops its own modeling strategy to identify partisan identification for every voter in a non-paid state. We use this information from the data provider as our baseline for party identification of the matched voters and their neighbors. 

In step two, we consider voters registered to a third party with a clear left or right ideological leaning as Democrats (left) or Republicans (right). Here, we follow closely the classifications scheme provided by \cite{brown2021measurement}. Table \ref{partylean} provides the classification of the third party's leaning. 

% code/table_party.r
\begin{table}[H]
\renewcommand{\arraystretch}{2}  % Adjusting the row spacing
\caption{\label{partylean}Classification Code for Third-Party Leaning}
\centering
\resizebox{\linewidth}{!}{
\begin{tabular}[t]{>{\raggedright\arraybackslash}p{2 cm}>{\raggedright\arraybackslash}p{15 cm}}
\toprule
Leaning & Third-Party Party\\
\midrule
Democrats & Democratic,Green Libertarian,Constitution,Green,Liberal,Progressive,Working Family Party,Peace And Freedom,Socialist,Socialist Labor,Rainbow,Bread And Roses,Worker's Party,Women's Equality Party,Social Democrat,Communist,Independent Democrat\\
Republican & Republican,Libertarian,Conservative,American Independent,Constitutional,Independent Republican\\
Unkown & Unknown,Non-Partisan,Registered Independent,Independence,Other,Natural Law,Reform,American,Peoples,Declined To State,Patriot,Consumer,Mountain\\
\bottomrule
\end{tabular}}
\end{table}

In step three, we impute the partisan identification for the remaining non-partisan voters  through a Bayesian process \citep{imai2016improving}, as described in the equation below:

\[
\text{Pr}(PID_{i} | X_i) = \frac{\text{Pr}(X_{i} | PID_i) \times \text{Pr}(PID_c)}
{\text{Pr}(X_{i} | R_i) \times \text{Pr}(R_c) + \text{Pr}(X_{i} | D_i) \times \text{Pr}(D_c) + \text{Pr}(X_{i} | I_i) \times \text{Pr}(I_c)}
\]

The Bayesian process contains two core parts: the individual level likelihood and precinct voter share prior information. The first component represents the likelihood derived from a set of demographics $X_i$. In this equation,  $X_{ij}$ represents the individual-level probability of a voter being part of a certain demographic group conditional on self-reported partisanship (Democrat, Republican,
or Independent). These probabilities are estimated using high-quality and large-scale survey data  from the Cooperative Congressional Election Study, and use all unique combinations between age groups (18-34, 35-50, 51-62, > 63), gender, and race.

These individual-level probabilities are derived directly from the survey data and are merged back into the profiles recorded for every remaining non-partisan matched voter and their neighbors in the voter files. However, for a small fraction of L2-matched voters, information about their gender and race is unavailable. For gender, we use information from the census data comparing first name and age on the frequency of male and female names across years to estimate the gender \citep{mullen2018gender}. To impute information about voter's race, we use a similar process relying on voters' last names and census information to estimate their racial group \citep{imai2016improving}. 

The prior component of the equation is constructed by calculating the precinct-level vote share of Presidential candidates in the 2020 election. Precinct-level vote shares were constructed from data provided by the nonpartisan Redistricting Data Hub (RDH) project, which contains both results and validated precinct boundaries shapefiles. Because the United States does not have a unified coding system for precincts, merging the voter files with precinct-level data requires a set of pre-processing steps. We use a three-step procedure to perform this task. First, we use exact textual matching between the precinct from the voter files and RDH precinct names. Second, we rely on fuzzy textual matching to perform the same task \citep{enamorado2019using}. Finally, for all remaining voters whose precincts we could not identify on RDH using textual matching, we rely on geospatial matching between their residential address and the RDH shapefile. This procedure allows us to identify the precinct for more than 98\% of all matched sample and their neighbors. 

After performing the steps described above, we estimate the quantities described in the imputation equation and impute partisan identity using the largest probability given by the Bayesian process. To assess the accuracy of our imputation, we augment the imputation to also include matched L2 voters for which we know their partisanship (as provided in step 1 for Democrats and Republicans). Our process achieves an accuracy score of 73.4\%, which is comparable to results reported in  \citep{brown2021measurement}.

\section{Robustness Checks: Discrete Assignment of Partisanship}

As discussed in the paper, we use the posterior probabilities from the imputation process to calculate the offline isolation measure. This is a more conservative measure in which we incorporate uncertainty from discrete choice models into the isolation measure. However,  rather than weighting neighbors by the probabilities, we can just assign voters to a partisan affiliation by using the highest probability calculated from the imputation process. In this section, we present results using the discrete assignment of the partisan identity. Results remain largely unchanged.

% code/descriptive_analysis_imputation_no_weights.r
%%% THIS FIGURES DO NOT EXIST YET
\begin{figure}[!htpb]
  \captionsetup{justification=raggedright, singlelinecheck=false}
  \caption*{\large\textbf{Probabilistic Assignment for Partisanship}}
\scalebox{.9}{\begin{tabular}{c}
\includegraphics[width=1\linewidth]{figs_imp/probs/noweights_imp_offline_isolation_quantiles_circles.pdf} \\
\end{tabular}}
  \captionsetup{justification=raggedright, singlelinecheck=false}
  \caption*{\large\textbf{Discrete Assignment for Partisanship}}
\scalebox{.9}{\begin{tabular}{c}
\includegraphics[width=1\linewidth]{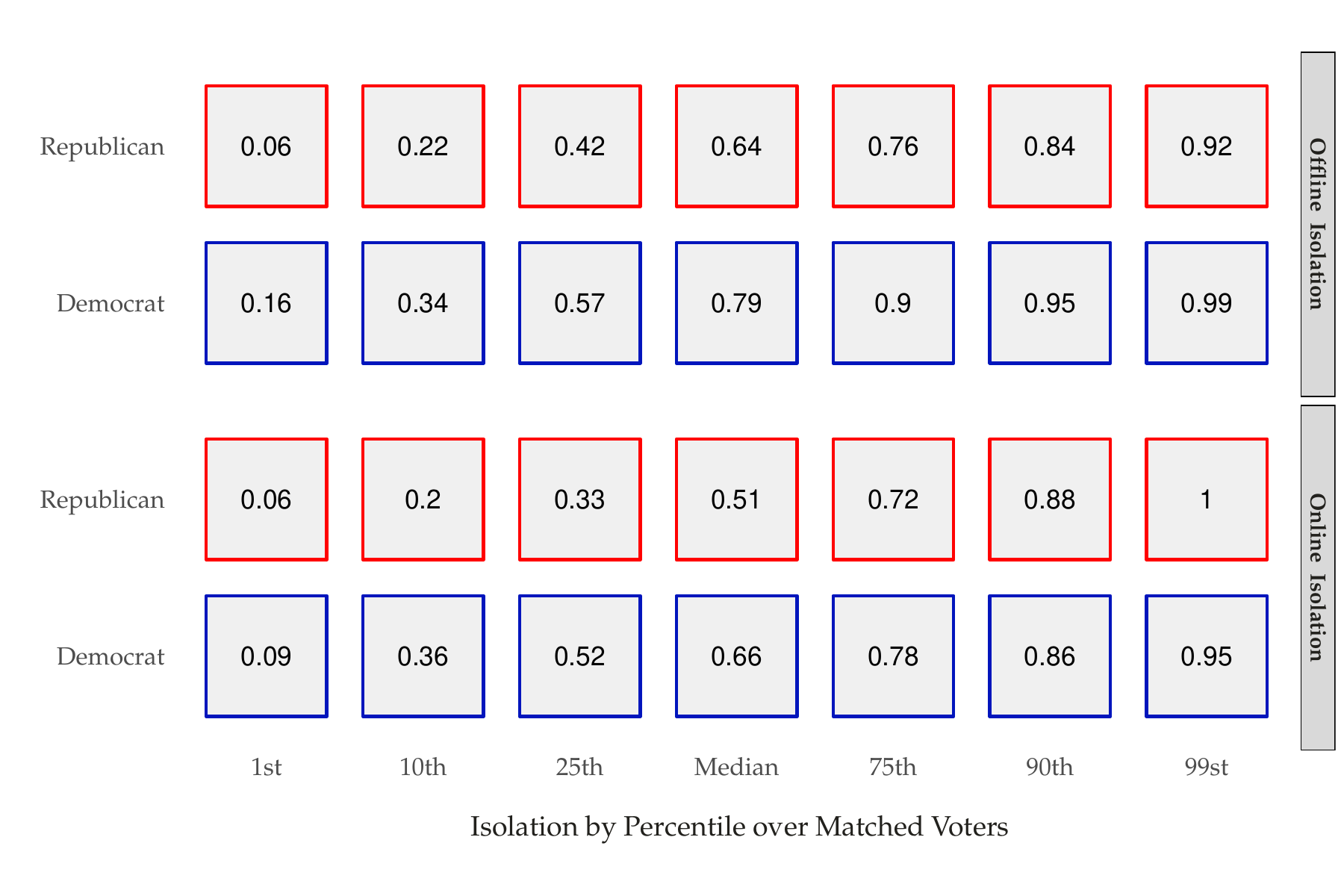} 
\end{tabular}}
\caption{\textbf{Partisan Offline and Online Isolation}. Democrats are shown in blue, and Republicans are in red.}
\label{fig:hist3}
\end{figure}

\section{Robustness Checks: Alternative Cutoffs for Online Isolation}

To estimate the partisan composition of online networks, we rely on ideal-points methods using as inputs the followers of friends of each matched voters in our panel. As described in \cite{barbera2015birds}, this method depends on having users that follow others for which we can have reliable measures of ideology. In our case, to calculate these scores, we use a set of elites, including members of Congress, prominent members of the executive branch (e.g., the President, the Vice President, and various cabinet members), and prominent political organizations (e.g., the RNC and DNC Twitter accounts). We then expand this list to a pre-curated list of 50 thousand twitter users that follow these accounts, and estimate the ideology score for these users. To estimate the ideology of the friends of our matched voters, we need these users to follow some of these political elites or the Twitter users we have already estimated the scores for previously.

\begin{figure}[!htbp]
\begin{tabular}{c}
\includegraphics[width=1\linewidth]{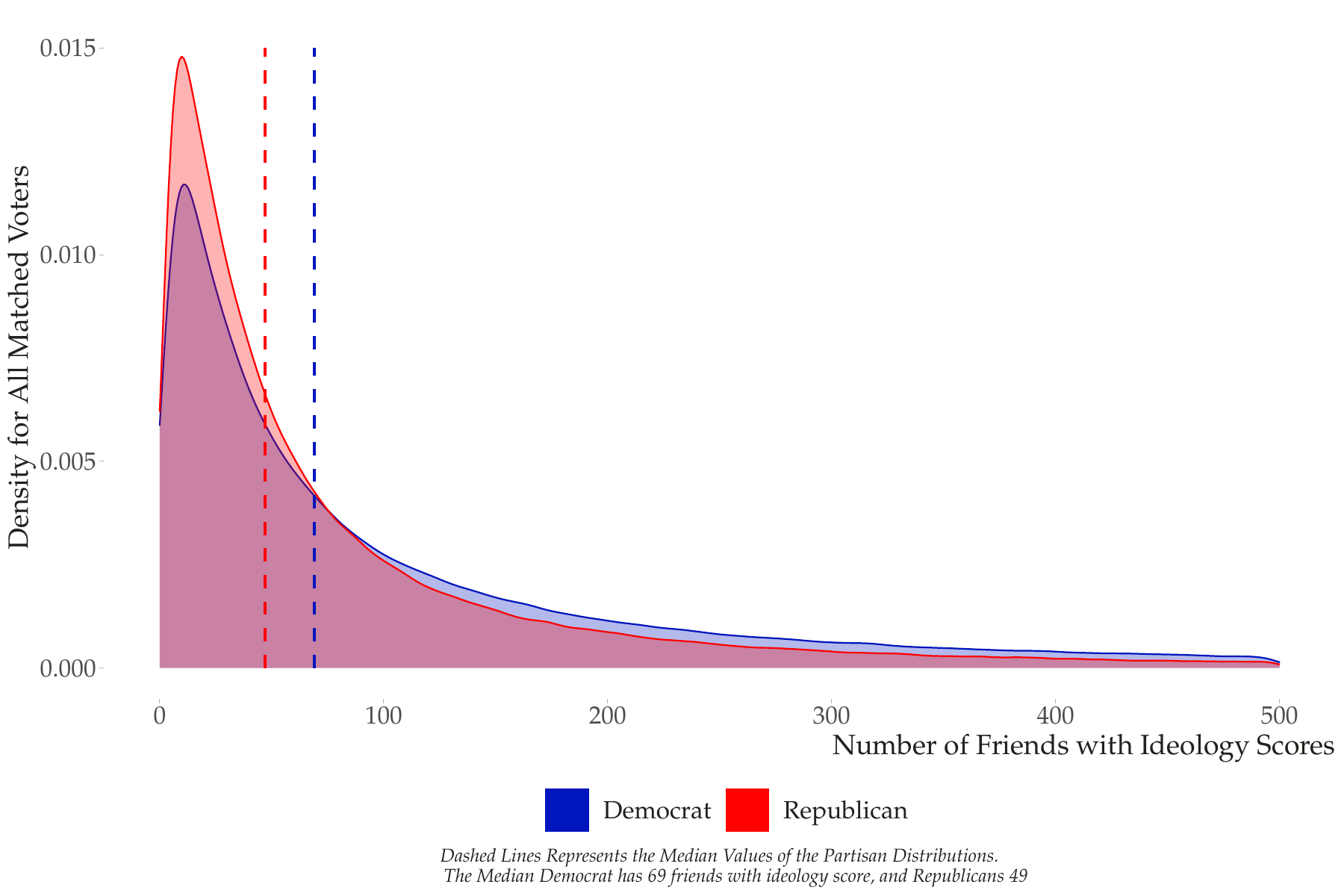}  \\
\end{tabular}
\caption{\textbf{Distribution for the Number of Friends with Ideology Scores}}
\label{sm_density}
\end{figure}

\begin{figure}[!htbp]
\begin{tabular}{c}
\includegraphics[width=1\linewidth]{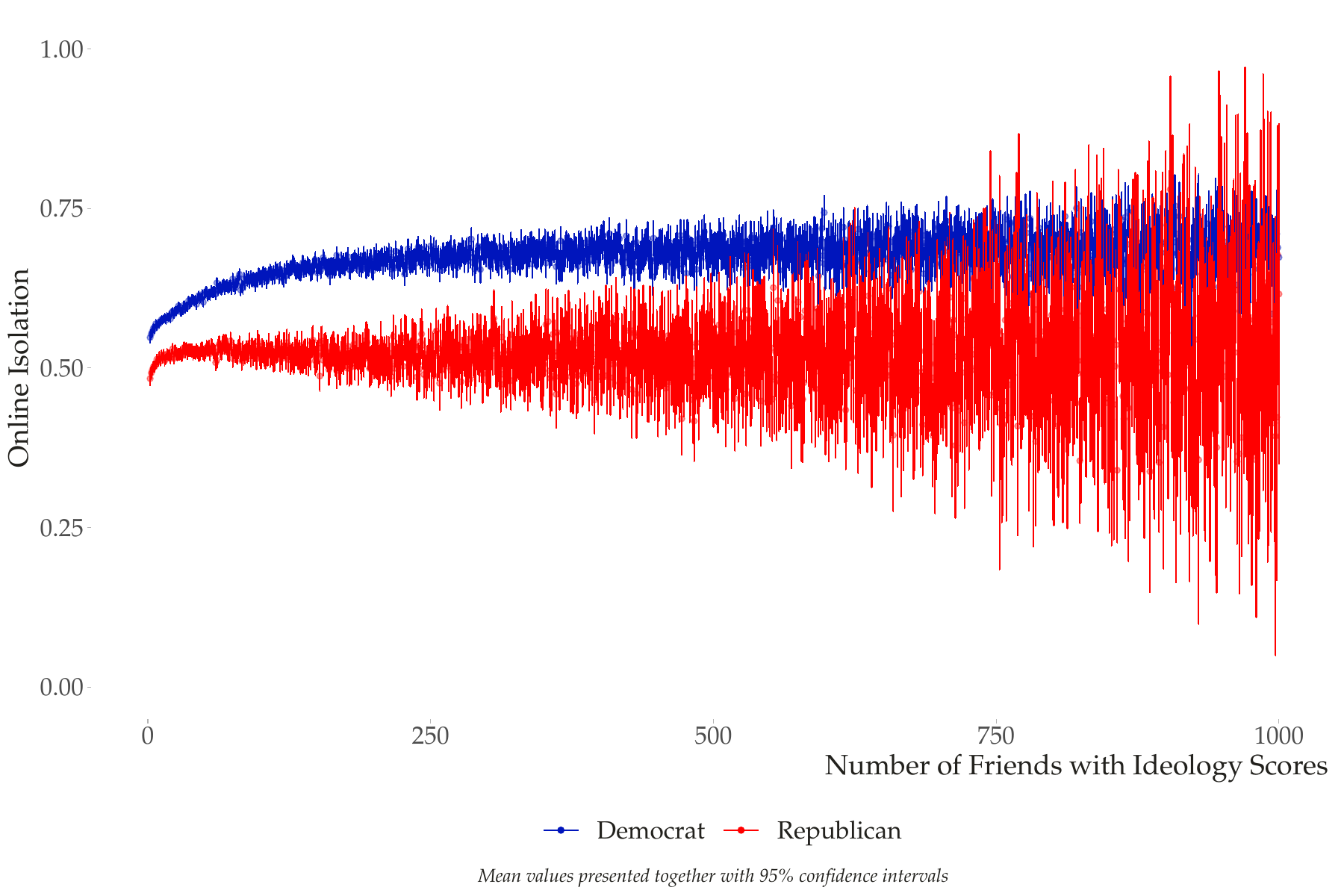}  \\
\end{tabular}
\caption{\textbf{Mean Online Isolation Across Users with Distinct Number of Friends with Ideology Scores}}
\label{sm_density_online}
\end{figure}

\begin{figure}[!htbp]
\begin{tabular}{c}
\includegraphics[width=1\linewidth]{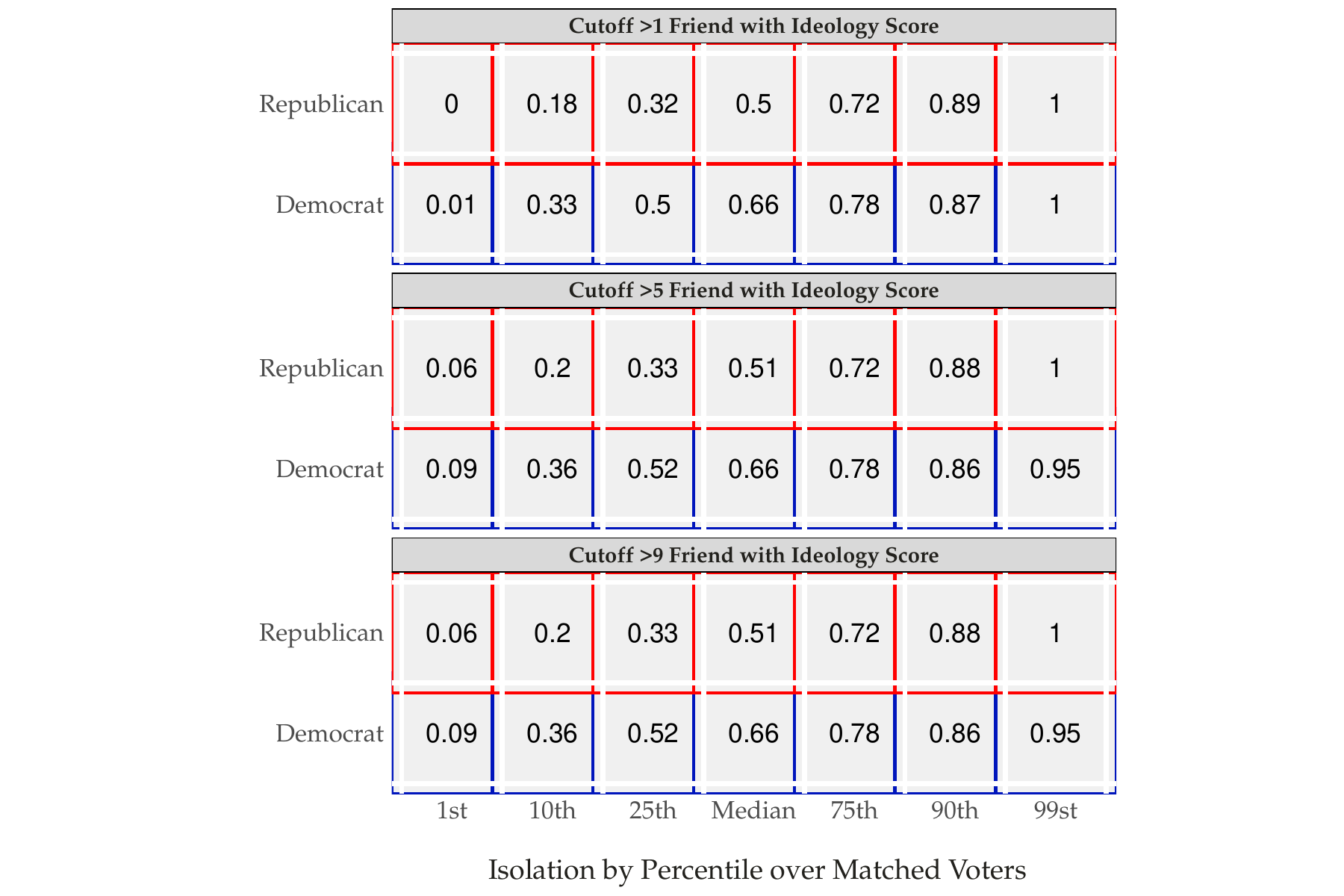}  \\
\end{tabular}
\caption{\textbf{Online Isolation Under Different Cutoffs}}
\label{sm_cutoff}
\end{figure}

On average, we can calculate ideology using this method for 30\% of the friends of our matched voters. As described in the paper, to make inferences about the online network of users based on very few online friends, we estimate online isolation measures only for those who follow at least 10 other Twitter users for which we can estimate an ideology score. In this section, we demonstrate the robustness of our results to this decision. In plot \ref{sm_density}, we show the distribution of the number of friends with ideology scores across the entire matched panel. In Figure \ref{sm_density_online}, we display the mean online isolation and 95\% confidence intervals for users with one to 1000 friends, for whom we can estimate ideology, illustrating that levels of online isolation do not vary conditionally on the number of friends with whom one has ideology. Lastly, in figure \ref{sm_cutoff}, we replicate Figure 2 in the paper using two alternative cutoffs to filter the data before estimating levels of online isolation. The latter plot shows median values of online isolation do not vary across the cutoff, but we remove extreme cases, with online isolation at zero or one, by filtering the data with a higher cutoff (>10 friends with ideology)

%\newpage

%\newpage

%\bibliography{bib}

%\end{document}

\end{document}